%% file: main.tex
\documentclass{iosart2c}
\usepackage[utf8]{inputenc}

\usepackage{times}%

\usepackage{graphicx}

\usepackage{rotating}

\usepackage{caption}

\usepackage{hyperref}

\usepackage{cleveref}

\usepackage[xindy, toc, acronym]{glossaries}

\usepackage{multirow}
\usepackage[table,xcdraw]{xcolor}

\usepackage{color}

\pubyear{2017}
\volume{0}
\firstpage{1}
\lastpage{1}

\begin{document}
\include{glossaire}

\begin{frontmatter}

\title{Capturing the contributions of the semantic web to the IoT: a unifying vision}
\runningtitle{A unifying vision for the Semantic Web of Things}

\address[A]{IRIT, UMR 5505, UT2J, 
D\'{e}partement de Math\'{e}matiques-Informatique, 
5 all\'{e}es Antonio Machado, \\
F-31058 Toulouse Cedex, France \\
E-mail: firstname.lastname@irit.fr}
\address[B]{LAAS-CNRS, \\
7 avenue du Colonel Roche, \\
F-31400 Toulouse, France\\
E-mail: firstname.lastname@laas.fr
}
\address[C]{Univ de Toulouse, INSA, LAAS, F-31400, Toulouse, France}
\author[A,B,C]{\fnms{Nicolas} \snm{Seydoux}},
\author[B,C]{\fnms{Khalil} \snm{Drira}}
\author[A]{\fnms{Nathalie} \snm{Hernandez}}
\author[B,C]{\fnms{Thierry} \snm{Monteil}}
\runningauthor{N. Seydoux et al.}

\begin{abstract}
The Internet of Things (IoT) is a technological topic with a very important societal impact. IoT application domains are various and include: smart cities, precision farming, smart factories, and smart buildings. The diversity of these application domains is the source of the very high technological heterogeneity in the IoT, leading to interoperability issues. The semantic web principles and technologies are more and more adopted as a solution to these interoperability issues, leading to the emergence of a new domain, the Semantic Web Of Things (SWoT). Scientific contributions to the SWoT are many, and the diversity of architectures in which they are expressed complicates comparison. To unify the presented architectures, we propose an architectural pattern, LMU-N. LMU-N provides a reading grid used to classify processes to which the SWoT community contributes, and to describe how the semantic web impacts the IoT. Then, the evolutions of the semantic web to adapt to the IoT constraints are described as well, in order to give a twofold view of the convergence between the IoT and the semantic web toward the SWoT.

\end{abstract}


\end{frontmatter}

\section{Introduction}

The internet has evolved toward a \textbf{globally connected environment}. The interconnection of devices, smartphones, services, etc, form what is called the \textbf{\gls{iot}}. A detailed definition of the \gls{iot} is provided in \cite{Szilagyi2016}, embracing the diversity of nature and purpose of the so-called Things. In this paper, the term \gls{iot} refers to the technologies and principles enabling the deployment of devices and services networks. \gls{iot} networks are based on very diverse technological stacks, for hardware, software and communications. The integration of traditional technologies of the web in the \gls{iot}, such as URIs for resource naming, HTTP connectivity, or REST interfaces, is called the \textbf{\gls{wot}}, defined by \cite{itu2012} as "A way to realize the IoT where (physical and virtual) things are connected and controlled through the World Wide Web". This definition is used as a reference by the W3C Web of Things Working Group\footnote{\url{https://www.w3.org/WoT/WG/}}. 

Indeed, the \gls{iot} and \gls{wot} "Things" are now an every day life reality for many people. To support this claim, one might consider the increasing number of smart cities including but not limited to Dublin (IR)\footnote{\url{http://smartdublin.ie/}}, Santander (ES)\footnote{\url{http://www.smartsantander.eu/}}, Milton Keynes (UK), San Francisco (US), New York (US), Yokohama (JA), that are being equipped in order to monitor their environment and to offer innovative services. It is estimated that by 2020, \textbf{26 billion devices} will be connected by \gls{m2m} technologies, compared to the 0.9 billion connected in 2009\cite{gartner2013}. The exponential multiplication of connected devices is correlated with a dramatic increase in the volume of exchanged data.

The growing integration of connected Things to human activities is partly driven by the wide scope of application domains covered by the \gls{iot}: environmental metering, transportation, home automation, e-health, agriculture, manufacturing...

However, the diversity of \gls{iot}-based applications is restricted by the approach of the industry so far, which has been oriented toward \textbf{vertical silos}, as \cite{Desai2015} points out. Proprietary systems are designed with a specific purpose, and on top of the devices the vendor also distributes the application serving the purpose. This approach cannot match the diversity of possible scenarios driven by user requirements, and raises \textbf{interoperability issues}. Users should be allowed to combine their connected devices in a personalized fashion, and an application developers should be able to deploy generic applications that adapt to the available devices, which is totally opposed to vertical integration. Furthermore, such customization of one's \gls{iot} network requires devices to be able to understand each other: they must be \textbf{semantically interoperable}. \cite{Gyrard2015} and \cite{Aissaoui2016}  make a distinction between syntactic interoperability, or the ability for systems to exchange content, and semantic interoperability, the ability to understand the exchanged messages. The extreme technological diversity\cite{eclipse2017}, as well as the multiplicity of standards (including but not limited to oneM2M\footnote{\url{http://www.onem2m.org/}},  OIC\footnote{\url{https://openconnectivity.org/}}, AllJoyn\footnote{\url{https://allseenalliance.org}}, LWM2M\footnote{\url{http://www.openmobilealliance.org}}), makes syntactic interoperability at a large scale a non-trivial issue. However, solving syntactic interoperability issues only is not sufficient to achieve the complete potential of the \gls{iot}, and it is not in the scope of this paper.

Semantic interoperability is an important enabler for the future development of the \gls{iot} \cite{Murdock2016}. Indeed, the core of the \gls{iot} is \gls{m2m} communication. For instance, the observations gathered by sensors are meant to be distributed to other devices, and not to be human readable without processing. Since no human can re-contextualize exchanges between devices, \textbf{messages should be understandable from a machine to another}. 

As \cite{Corcho2010} or \cite{Murdock2016} points out, semantic web principles and technologies can provide solutions to the issues the \gls{iot} is facing. The use of dereferencable vocabularies such as ontologies enables the capture of metadata in a machine-understandable way, easing \gls{m2m} communication. Many recent research contributions from the semantic web community inject semantic web capabilities (rich content description, reasoning...) into the \gls{wot} in order to develop the so-called \gls{swot}, an evolution of the \gls{wot} where the \gls{iot} is extended with semantic web principles and technologies. The publication of resources from an \gls{iot} network to the \gls{swot} can feed the \gls{lod}, in order to make the collected data available.

This paper depicts a landscape of the \gls{swot}, and provides a reading grid to classify the research contributions transforming the \gls{iot} into the \gls{swot}. First, to structure the analysis, section \ref{sec:lmu} defines \gls{lmu}, a node-centric \gls{iot} architectural pattern. \gls{lmu} provides a framework to support the identification of recurrent patterns in existing research. It is used to describe how the semantic web principles and technologies contribute to the \gls{iot} in section \ref{sec:contributions}. On the other hand, section \ref{sec:transformation} focuses the challenges the semantic web faces to be compliant with the \gls{iot} constraints, and how recent semantic web contributions propose to adapt to them. Finally, section \ref{sec:conclusion} concludes this paper and proposes some perspectives for the future of \gls{swot}.

\section{Unifying the heterogeneity of architectures with LMU-N}
\label{sec:lmu}

After having motivated the need for a unifying architectural pattern, this section presents \gls{lmu} and its main components: nodes and messages flows. Other surveys are listed as related work, before showing how \gls{lmu} can be used as a reading grid for our survey.

\subsection{Motivations}

The papers associating the semantic web and the \gls{iot} are many, and their publication rate is increasing: from 15 publications in 2003, to 494 publications in 2016\footnote{After a study on \url{http://ieeexplore.ieee.org}, \url{http://www.sciencedirect.com} and \url{http://dl.acm.org}, searching for the exact keywords "semantic web" and "internet of things"}. The total count of publications amounts to 1426. We focused on 71 scientific publications for the survey, chosen for their quality, their innovative aspect, and for the balance in their content between semantic web and \gls{iot}. An extra attention was given to publications proposing semantic web contributions in explicitly and precisely defined \gls{iot} architectures. These publications describe contributions that can be competing or complementary, mutually exclusive or not, etc. In order to be able to capture a structured landscape of the contributions of the semantic web to the \gls{iot}, one needs to be able to compare these contributions, and to organize them with respect to one another. To do so, we propose \textbf{\gls{lmu}, a unifying architectural pattern} that aims at identifying the core components of an \gls{iot} network in a generic manner, and to characterize their relationship. \gls{lmu} is then used as a framework to contextualize semantic web contributions to the \gls{iot}. 

\subsection{Constituting elements of LMU-N}

\begin{figure}
\centering
\includegraphics[width=0.45\textwidth]{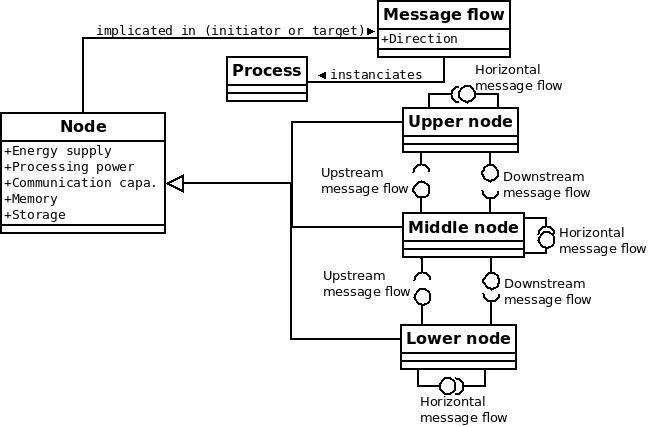}
\caption{The LMU-N pattern}
\label{fig:lmu_overview}
\end{figure}

\gls{lmu} has two major components: the \textbf{nodes}, or the "Things" communicating on the \gls{iot}, and the \textbf{flows}, representing the communications between the nodes. \gls{lmu} is the result of a bottom-up analysis, and its description is linked to a set of pre-existing architectures from which it was deduced. The pattern is represented as an UML model in figure \ref{fig:lmu_overview}. The pattern is composed of three main classes: the Node, specialized in three subclasses, the Message flow, and the Process. The nature of these classes are respectively described in sections \ref{subsub:node}, \ref{subsubs:flows} and \ref{subsubs:processes}.

\subsubsection{Representing physical device and virtual service as Nodes}
\label{subsub:node}

\paragraph{Definition of an \gls{iot} node}

The \gls{iot} is a network of Things connected together, and can be seen as a \textbf{graph where the Things are vertexes and their connections are the edges}. So far, the notion of Thing has been used to indifferently refer to devices (sensors and actuators) and to services. This choice is justified by the similarity of the roles devices and services have on \gls{iot} networks: a service can be seen as the endpoint of a device for the other nodes of the \gls{iot}, and a device can be seen as the physical implementation of a service, depending on the chosen perspective.

That is why we introduce in this section the \textbf{node}, an abstraction covering both device and service concepts. Fundamentally, \textbf{a node is an active entity that can be addressed on the network}. An active entity is able to send and/or receive requests, which is the case for both devices and services.  This notion of node is already present in architectures described in \cite{Alaya2015} or in \cite{Perera2014_context}. \cite{Alaya2015} describes a contribution based on oneM2M\footnote{\url{http://onem2m.org/}}, a standard architecture structured around different nodes which interfaces are formally defined to enforce interoperability. Blurring the line between physical object and virtual service is also at the core of propositions such as \cite{Pfisterer2011}, \cite{Foteinos2013} or \cite{Kibria2015}, where "virtual entity" is used as an abstraction to enable the composition of services or devices.

Section \ref{sec:contributions} lists approaches that are either directed to devices or services, and their seamless integration and comparison is enabled by the adoption of the node as the central architectural element. Moreover, presenting nodes over devices or services serves the purpose of interoperability by easing the homogeneous modelling of their capabilities. For instance, virtual entities in \cite{Mrissa2015} are used to capture the characteristics of a node making it relevant to its neighbours. 

\textbf{Abstracting devices and services} into a generic node also aims at \textbf{focusing on their intrinsic characteristics}, such as processing power or communication abilities. Since the early studies on semantic sensor networks, the description of devices and services has been a primary concern of the \gls{swot} community. Contributions like \cite{MohammedEid7}, \cite{Lefort2011} or \cite{Seydoux2016} propose ontologies to define both the capabilities of a device, and to add metadata to the content it handles. Section \ref{subssubs:node_processes} details how node descriptions are used to integrate an \gls{iot} deployment in the \gls{swot}.

\paragraph{Classification of \gls{lmu} nodes}

\textbf{All nodes of an \gls{iot} are not equivalent}: some devices are very constrained, whereas the servers providing analytic capabilities are powerful machines. For instance, in its vision of urban \gls{iot}, \cite{Zanella2014} uses the notion of node, and proposes to use two different protocol stacks according to the capabilities of each node (constrained or not). In this paper, different characteristics are used to cluster nodes into meaningful classes constituting \gls{lmu}, represented as attributes of the Node class in fig. \ref{fig:lmu_overview}. The core characteristic of a node is its \textbf{processing power}, i.e. its ability to apply treatments of varying complexity to content. The processing power also determines the ability of the node to process content of a varying expressivity, from the very simple agreed-upon byte array to the much more complex \gls{kb} instantiations. The higher a node's processing power is, the more expressive content it can process, and the more complex operations it can achieve. Nodes are also characterized by their \textbf{memory}, i.e. the quantity of information it can hold at a given time, and \textbf{storage} capability. Storage is the available space giving access to persistent content. The notion of \gls{iot} is inseparable from the notion of connectivity, and a node can also be classified according to its \textbf{communication capabilities}. These capabilities include the protocols it supports, its general availability on the network, and its bandwidth. Nodes also differ by the nature of their energy source: while some nodes are attached to traditional power grids, other nodes, deployed in the field, are reliant on batteries, or on renewable energy sources like solar panel, or energy harvesting. 

Three homogeneous types of nodes are identified, constituting the \textbf{three layers} of \gls{lmu}. These subclasses of Node are represented on fig. \ref{fig:lmu_overview}, and are described in tables \ref{table-node-roles} and \ref{table-node-caracs} regarding their typical characteristics and roles on the network.

\begin{itemize}
	\item \textbf{\gls{un}} are the "weakly constrained nodes" of \cite{Zanella2014}, or Infrastructure Node in \cite{Alaya2015}. In this category, we classify cloud or local servers as well as powerful mobile devices, including standard laptops and smartphones and domain-specific mobile robots or machines. They have high processing power, extended communication capabilities, and large storage capabilities. For instance, in \cite{DahnLe-Phuoc18}, \gls{un} are in charge of applying data fusion operators to data streams. In \cite{Liu2015} or \cite{Szilagyi2016}, the \gls{un} are situated in the cloud. 
	\item \textbf{\gls{mn}} are very often referred to in the literature as \textbf{gateway} (\cite{Compton2009}, \cite{Alaya2015}, \cite{Desai2015}), because they are bridges between powerful \gls{un} and more constrained nodes. They are usually dedicated to content transformation and protocol bridging: in \cite{PayamBarnaghi118}, \gls{mn} are presented as intelligent nodes where content can be converted from its raw representation to a richer one. \cite{Desai2015} proposes an architecture where \textbf{the gateway is a contact point between the \gls{iot} and the \gls{swot}}, performing both annotation and protocol proxying. In \cite{Nikoli2011} and \cite{Zanella2014}, \gls{mn} are proxies for wireless devices networks. \gls{mn} typically have medium processing power, extended communication capabilities, and restricted memory storage: they are nodes where content collected by sensors or sensing services is collected, transformed, and redistributed. 
	\item \textbf{\gls{ln}} are typically connected devices, with very limited power source, processing and communication capabilities, and very limited to no storage capabilities. These are by definition present in every \gls{iot} architecture, and in direct contact with the physical world. In some papers such as \cite{DahnLe-Phuoc18} or \cite{Vlacheas2013}, these nodes are not directly present, their representation is wrapped by a \gls{mn}. The \gls{ln} were in early studies mainly sensors, like in \cite{Compton2009}, and evolved toward diverse nodes including actuators, displays and composite devices in more recent work such as \cite{Alaya2015}.
\end{itemize}

\begin{figure}
\centering
\includegraphics[width=0.45\textwidth]{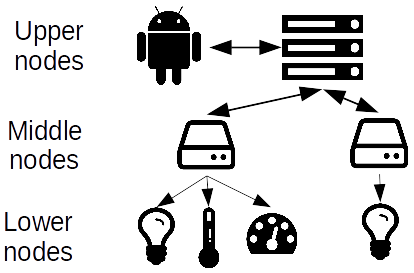}
\caption{An instantiation of LMU-N}
\label{fig:pattern-instantiation}
\end{figure}

Figure \ref{fig:pattern-instantiation} represents an instantiation of the \gls{lmu} pattern based on the work described in \cite{Seydoux2016}. Two lamps, a temperature sensor, and a pressure sensors instantiate lower nodes. Their data is collected in a request-response fashion by two gateways. The communication is initiated by the gateway, therefore the arrows are directed from the gateways to the devices. The two upper nodes represent the server hosting the enrichment application, and the robot which processes the content generated by the sensors. The gateways themselves have push capabilities, and can also be queried directly by the server: the communication goes both ways. SmartSantader\footnote{\url{http://www.smartsantander.eu/}} gives another simple example of real-world deployment conform to the \gls{lmu} pattern.

\input{tables/table_node_caracs}
\input{tables/table_node_roles}

All the architectures presented in this section can be described by \gls{lmu}, supporting its unifying status. A notable exception to the three-layers architecture model is presented in \cite{Hasemann2012}, where the authors promote a direct connection between devices and application servers. They propose to this end an architectural approach where each node is equipped with a high-efficiency RDF serializer and tuple store. However, in our opinion, this approach requires the devices to be able to run such a stack, and reduces the possibility to include non semantic-aware, legacy devices in an \gls{iot} system. Moreover, we think that directly connecting \gls{ln} to \gls{un} in a flat architecture is a source of scalability issues in large systems. That is why \gls{lmu} is structured with an intermediary level to mediate between potentially very constrained nodes and the rest of the network.

\subsubsection{Representing exchanges between nodes as messages flows}
\label{subsubs:flows}

In \gls{lmu}'s representation of an \gls{iot} network as a graph, the edges are communications between nodes, instantiated by messages flows. We will refer to these messages' expressivity according to the classification provided by \cite{Rowley2007} as data, information or knowledge, organized in the \gls{dikw} hierarchy. Typically, \textbf{sensors produce data}, and this data is later on contextualized to become information, and enriched using semantic web principles and logic in order to produce knowledge.  

Furthermore, the \gls{iot} relies on already existing internet technologies and paradigms. Nodes hosting services are servers, and they offer their services to applications or other nodes that are clients in a classic client-server architecture. This model implies asymmetry in the communications among nodes: in a client-server model, the interaction is initiated by the client. This entails that on the graph of nodes, \textbf{the edges are oriented}, from the client to the server node. 

As shown on fig. \ref{fig:lmu_overview}, flows can be directed in three general directions: \textbf{horizontal} for nodes of the same level, such as the work done in \cite{Ma2014}, \textbf{upstream} (as in \cite{AmitSheth30} or \cite{CoryHenson115}), when the source node is of a lower level than its destination node, and \textbf{downstream} (as in \cite{Poslad2015}) otherwise. The notions of upstream and downstream communication in \gls{iot} architectures are presented by \cite{Poslad2015}. The content exchanged in these flows can either be application-dedicated, for instance in \cite{DahnLe-Phuoc18}, \cite{CoryHenson115} or \cite{Poslad2015}, specific to the function of the devices that collected it (temperature observations, user requests...), or system-dedicated, describing the nodes constituting the network, like in \cite{Nikoli2011}. 

\subsection{Using LMU-N to classify \gls{swot} contributions}
\label{subsubs:processes}

By describing \gls{iot} systems with the \gls{lmu} pattern, \textbf{different recurring processes are identified, agnostic to the underlying technology stack and to the application domain}. The aim of this paper is to identify and describe the processes that can benefit from an integration into the \gls{swot}, and how they are instantiated in research contributions. 

The processes are instantiated by exchanges of messages between nodes, supported by flows as described in section \ref{subsubs:flows}. Therefore, our analysis is based on the assumption that \textbf{(i) processes are directed} (horizontal, upstream or downstream) and \textbf{(ii) dependant on the characteristics of the nodes at stake}. Contributions brought to the \gls{swot} are dedicated to certain processes, and they can be located in \gls{lmu} depending on the nodes they involve. For instance in \cite{Ming2013} or \cite{Le-Phuoc2016}, the process of content enrichment is an upstream process, as the content is produced in lower nodes and enriched later on in middle or upper nodes. Processes can also be associated with specific flows content: for instance, in \cite{Ruta2016}, the node exposition process is based on node metadata, and not with applicative content.

All the processes introduced in this paper are identified from published papers after their projection on \gls{lmu}. Two categories of processes are distinguished, presented in the remainder of this section.

\subsubsection{Content-related processes} 

Content-related processes are the activities that are specific to an application domain, and that focus on the \textbf{transformation}, \textbf{transport} and \textbf{processing} of content relevant to an application. Tables \ref{lmu-table-content_1-2} and \ref{lmu-table-content_2-2} classify papers contributing to these processes, and situate them into \gls{lmu}. Section \ref{subs:content-process} details the processes of this category.

\subsubsection{Node-related processes}
\label{subssubs:node_processes}

Node-related processes are activities not specific to an application domain, but common to the \gls{iot} domain. The development of such processes goes against vertical fracturation. The messages exchanged in these processes focus on the description of nodes characteristics, capabilities, preferences, etc, in order to allow nodes to be \textbf{aware} of their neighbours on the graph, and to offer a \textbf{homogeneous} representation. Table \ref{lmu-table-node} references the papers contributing to these processes, and they are described in section \ref{subs:node-process}.

\subsubsection{Ontologies of the \gls{swot}}

In order to integrate \gls{iot} deployments in the \gls{swot}, their resources must be described using ontologies. Depending on the type of process, the ontologies required for the \gls{swot} integration is different. 

In the case of content-related processes, there is a need for ontologies capturing domain knowledge. The LOV4IoT\footnote{\url{http://www.sensormeasurement.appspot.com/?p=ontologies}} initiative mimics the principle of the LOV\footnote{\url{http://lov.okfn.org/dataset/lov/}}, and references ontologies that have a connection with the \gls{iot} domain, either describing \gls{iot} concepts, or domain knowledge in a field impacted by the \gls{iot}, e.g. agriculture, home automation, or e-health. These ontologies are diverse, and disconnected from each other: the vertical fracturing between domains is not bridged. 

Node-related processes are more focused on interoperability, and require the modelling of \gls{iot} systems common characteristics despite their diverse domains of interest. \gls{iot} ontologies capturing these connected device network characteristics exist. Several ontologies have been proposed in order to describe nodes: \gls{ssn}\footnote{\url{http://purl.oclc.org/NET/ssnx/ssn}}, saref\footnote{\url{https://w3id/saref}}\cite{Daniele2016}, iot-ontology \footnote{\url{http://ai-group.ds.unipi.gr/kotis/ontologies/IoT-ontology}}, IoT-lite \footnote{\url{http://iot.ee.surrey.ac.uk/fiware/ontologies/iot-lite}}\cite{Bermudez-Edo2016}, Spitfire \footnote{\url{http://sensormeasurement.appspot.com/ont/sensor/spitfire.owl}}, IoT-S\footnote{\url{http://personal.ee.surrey.ac.uk/Personal/P.Barnaghi/ontology/OWL-IoT-S.owl}}, SA\footnote{\url{http://sensormeasurement.appspot.com/ont/sensor/hachem\_onto.owl}}, the oneM2M base ontology\footnote{\url{http://www.onem2m.org/ontology/Base\_Ontology/}} and IoT-O \footnote{\url{http://www.irit.fr/recherches/MELODI/ontologies/IoT-O}}.
Papers like \cite{Compton2009}, prior to the apparition of SSN, or \cite{Szilagyi2016}, much more recent, survey these ontologies.

In both cases, the diversity of ontologies at stake can be an issue. That is why good practices must also be followed in the conception and the integration of these ontologies, described for instance in \cite{Gyrard2015} or \cite{Seydoux2016}.\newline

To sum up, \gls{lmu} is an architectural pattern built from a bottom-up study of \gls{iot} architectures. It is composed of nodes and messages flows, two abstractions that can be described using ontologies. \gls{lmu} is not the core contribution of this paper, but it provides a reading grid that is used in the next section. It is a relevant classification framework: 72\%
of the contributions listed in section \ref{sec:contributions} can be at least partially situated within \gls{lmu}.

\subsection{Related work}
Previous work has been done to survey the convergence between the \gls{iot} and the semantic web: 
\begin{itemize}
    \item Early work in the \gls{swot} focused on semantic sensor networks. For instance, \cite{Compton2009} surveys sensor ontologies and observation representations. The scope of this work is especially on models, and even if it proposes an overview of technologies enabling semantic sensor networks, it does not present specific applications. Similarly, \cite{Szilagyi2016} gives an overview of the semantic web stack applied to the \gls{iot}, and surveys \gls{iot} ontologies. It goes beyond semantic sensor networks, but is still limited to models analysis. We propose in this paper to focus on how the ontologies and the technologies of the semantic web are used to develop the \gls{swot}, rather than on identifying exhaustively the models used.
    \item \cite{Atzori2010} is a survey of the \gls{iot} domain, proposing a definition for the notion of \gls{iot} and listing application domains and enabling technologies for the \gls{iot}. The \gls{iot} paradigm is described as the convergence of Internet technologies, electronic devices, and semantic web technologies. However, the paper itself does not cover how semantic web technologies are integrated into the \gls{wot}, while we intend to analyze in detail and compare different contributions to the \gls{swot}.
    \item \cite{PayamBarnaghi117} studies the roles semantic web technologies can play in the \gls{iot}, as well as the challenges they represent. This paper identifies some processes similar to what is presented in this paper in section \ref{sec:contributions}. However, we situated the contributions to these processes within \gls{lmu} in order to enable a deeper understanding of how an \gls{iot} network is integrated into the \gls{swot}, and to give a finer-grained analysis grid: we want to identify the impact of the nodes on the processes they are involved into. The explicit identification of processes, as well as the characterization of nodes based on identified criteria, also \textbf{allows future studies to be classified} within the analysis grid we propose.
    \item \cite{Jara2014} gives an overview of the evolution from the \gls{iot} to the \gls{wot} and toward the \gls{swot}. It is focused on the role of standards in interoperability, and the integration of semantic web technologies in standards. It also provides an overview of technologies at stake in the \gls{iot}. This paper is oriented toward projects and industrial consortiums, which is out of the scope of our study. We focus on the contributions of the \gls{swot} to \gls{iot} issues,  and only integrate standardization concerns when they are related to this domain. 
\end{itemize}  

\section{Contributions of the semantic web to IoT processes}
\label{sec:contributions}

In this section, we describe for each process the contributions of associated papers to point out the added value of semantic web principles and technologies. Each category of process (content-related and node-related) are separated into subcategories of general recurring patterns among processes in order to cluster them into coherent sets: complementary processes, or processes implicated in similar scenarios.

The complete classification is summarized in tables \ref{lmu-table-content_1-2}, \ref{lmu-table-content_2-2} and \ref{lmu-table-node}. In these tables, each column represents a process, and each line represents the situation of the contribution in \gls{lmu}: if applicable, the contribution is precisely situated between a source and a target node type. However, due to the lack of information in some paper, the positioning of some contributions is only partial (in the lines "unspecified"). Moreover, some contributions are described completely decorrelated from any architecture, or in an architecture not compatible with \gls{lmu}, they are placed in the "Non-LMU/Not specified" line.

\subsection{Content-related processes}
\label{subs:content-process}

Content-related processes classification is summarized in tables \ref{lmu-table-content_1-2} and \ref{lmu-table-content_2-2}. Three main categories of content-related processes are distinguished, shown in the headers of the tables: representation transformation, transport, and processing, each described in the remainder of this section.

\subsubsection{Content representation transformation}

These processes are dedicated to the transformation of content according to the \gls{dikw} hierarchy. \textbf{The core meaning of the content is not changed} by these processes, but the expressivity of its representation varies.

\paragraph{Enrichment} is the process of transforming content representation upward in the \gls{dikw} hierarchy. For instance, the description of data with meta-data to transform it into an information is an enrichment: content is described with ontological entities to unambiguously define its meaning, capture its context and increase its reusability and its overall value to an application. Enrichment is one of the earliest processes that used semantic web technologies to contribute to the \gls{iot} with contributions such as \cite{MohammedEid7} or \cite{AmitSheth30}. This matter of fact is explained by the predominance in early \gls{iot} work of sensor networks, a subset of \gls{iot} networks where all the \gls{ln} are sensors. Enrichment is performed when content is exchanged upward in \gls{lmu}: all the enrichment contributions in table 3 are directed upward, which makes enrichment an upstream process, increasing content complexity and processing cost.

In \cite{AmitSheth30}, the enrichment is performed by \textbf{annotating the content} with RDFa\footnote{\url{https://www.w3.org/TR/xhtml-rdfa-primer/}}. This approach does create semantically enriched content, but it requires the \gls{ln} to produce complex, structured content (here compliant with the OGC's O\&M model\footnote{\url{http://www.opengeospatial.org/standards/om}}). This might be impossible for some \gls{ln}, as it requires to exchange heavy XML documents over communication links that can be very restricted (LPWAN networks or CoAP for instance). To overcome this issue, \cite{PayamBarnaghi118} proposes to \textbf{push the enrichment task to the \gls{mn} level}: the \gls{ln} produces raw data, in any format, this data is transported to the \gls{mn} by a dedicated network, and only then the \gls{mn} uses its own knowledge of the producing node to enrich the data. \cite{Kiljander2014} instantiates enrichment at the same level as \cite{PayamBarnaghi118} (as shown on table \ref{lmu-table-content_1-2}), from \gls{ln} to \gls{mn}. However, contrary to \cite{PayamBarnaghi118}, \cite{Kiljander2014} proposes an approach where the \gls{ln} offers a "semantic interface", implementing enrichment functionalities: content is enriched on the \gls{ln} side. Unfortunately, the paper does not give any additional precision to describe its enrichment approach. In a different approach, enrichment to a semantic format is performed in \cite{Aissaoui2016} and \cite{Seydoux2016} on the \gls{un} side when data is collected from the \gls{mn}. Content collected from sensors is stored in a standardized structure on the gateway, but it is explicitly described with an ontology only on the server side. 

The importance of the enrichment process is underlined by the important number of publications contributing to it: 17 papers in this survey. It comes from the discrepancy between the \gls{ln}, hardly able to produce high-level content, and the \gls{un} having applicative need for homogeneous content represented in expressive formats. Moreover, transforming data into knowledge is a common practice in the semantic web domain, even outside the \gls{iot}. This is an explanation to the fact that 47\% of the contributions to this process cannot be situated precisely into \gls{lmu}: their approaches are more centered on content than on the architecture around it.

\paragraph{Lowering} is the process exactly \textbf{opposed to enrichment}: it is the transformation of content from a semantically rich format to a less expressive, more constrained representation. The need for such transformation arose from the integration not only of sensors, but also of actuators as \gls{ln} in the \gls{iot}. These devices are content consumers, but their constrained nature prevents them from being able to consume some type of content. Actions represented in rich formats by \gls{un} need to be adapted to the target device so that it can interpret it and act as intended. This transformation deprives the content from part of its expressivity and context, but the transformed, simpler content is interpreted in a known context, leading to a \textbf{trade-off for consistency}. Lowering requires a node to have a representation of the capabilities and expectations of a remote node: it is a process that needs the source node to be more powerful than the destination node, and that is necessary because of the restrictions of the destination node. It is therefore a downstream process.

Lowering is opposed in the literature to "lifting", a synonym for enrichment. \cite{Kopecky2007} describes SAWDSL, a language aiming at making it possible to lift XML to RDF and to lower RDF to XML thanks to XML schema annotations. However, in the studies surveyed here, most of the approaches to lowering are not explicitly based on annotations, but rather on ad-hoc approaches. For instance, in \cite{Seydoux2016}, an autonomic control loop is instantiated, and high-level representation of actions are transformed into service calls, but the transformation process ad-hoc: RDF individuals and their OWL descriptions are consumed by a software that call the procedures based on an a priori interpretation of the expressed content.

Overall, the transformation of content from a high-level representation to a form that can be processed by a constrained node is still a challenge for the \gls{swot}. In the publications referenced in this paper, the proposed mechanisms are either \textbf{manual or ad-hoc}.  A majority of the studies cannot be clearly situated in \gls{lmu}, denoting shallow contributions. The lack of interest in this process comes from its contradiction with the usual practices in the semantic web community. Content is generally moved upward in the \gls{dikw} pyramid, because it gains value this way. However, the presence of \textbf{constrained content consumers} (actuator nodes) on the \gls{iot}, combined to the need for high-level content in \gls{un}, makes lowering a process as necessary as enrichment for a real \gls{swot}.

\subsubsection{Transport}

In processes listed under the transport category, content itself is not transformed. These contributions focus on how content is distributed across an \gls{iot} network (e.g. routing), or on how the content is used (e.g. control). Content itself is not necessarily expressed using the semantic web technologies, but its management on the network is based on these technologies.

\paragraph{Notification/Dissemination} is the process of forwarding a piece of content to a distant node \textbf{in a push manner} (initiated by the content collector). It can be driven by the interest of the receiver, or on applicative logic of the emitter. Content is disseminated from the node that generated it, either by collection (typically by a \gls{ln}), or by a higher-lever process (e.g. enrichment by a \gls{mn} or processing performed by a \gls{un}) toward another node. The diversity of nodes being sources or targets for this process makes it, in theory, a mixed process, going both upstream from \gls{ln} and downstream from \gls{un}. This mixed aspect is present for instance in \cite{Poslad2015}: on the one hand, sensor data is sent upstream, toward a control center. The control center processes the data, and produces alert messages that are disseminated downstream. 

\cite{Qin2015} proposes a dissemination approach based on a ad-hoc interest expression for content consumers represented by a query. Each consumer query is registered on a content matching component in the form of predefined triple pattern. Triples matching a registered pattern will be redirected to the relevant content consumer. However, this approach is oriented toward performance, and reduces the expressivity of queries, limited to a unique triple where each element can be fixed to a constant value. Approaches based on stream processing, such as \cite{Le-Phuoc2016} or \cite{Calbimonte2011}, use the full expressivity of SPARQL coupled to specific streaming operators, based on a time window or a number of events. The result of a stream query is an event stream, which is directed toward the querying node.

Even if fundamentally, dissemination is a mixed process, in the surveyed papers notifications are mainly sent upstream. This can be related to the architectural tendency of \gls{iot} networks to a hierarchical organization, with complex processings performed in the upper nodes. Therefore, it makes sense that \gls{un}, driven by applicative needs, express their needs to nodes, lower than them, that will collect content in a general purpose approach. The absence of goal-awareness in content collection by \gls{ln} is motivated by the will to \textbf{avoid stovepiped deployments}.

\paragraph{Control} is the process where a node sends a command to a remote node for it to execute. Commands are either issued by the user, or by high-level applications executed on \gls{un}, and are eventually executed by a \gls{ln}, making control a downstream process. In this process, the use of semantic web technologies and principles allows the target node to have a deeper understanding of the command.

\cite{IoannisChatzigiannakis129} propose an approach based on rule evaluation: if certain conditions (expressed in RDF) are matched, the rule entails an action. This approach allows the same node to have different behaviours upon the reception on the command, if the rule head relies on agents contextual knowledge. In \cite{Seydoux2016} and \cite{Aissaoui2016}, a representation of the actions to be taken is inferred from the representation of the user requirements and from the observations of the environment. These actions are associated to service description that are processed by a remote node to actually make the service calls. 

The necessity for the control process arises from the integration into \gls{iot} networks of \textbf{actionable nodes}. The reduced number of actionable nodes deployed compared to sensors explains that only three papers discuss contributions to this process. 

\paragraph{Routing} is the process where a route among nodes is determined for a content instance. Routing policies are at the core of the internet, and not only of the \gls{iot}. The integration of the semantic web technologies enables the definition of new routing policies. Routing policies are not specifically directed toward a direction in \gls{lmu}, therefore routing is a mixed process. Routing is different than dissemination in the sense that it is implemented by nodes on content they did not necessarily generate. Moreover, routing tends to be driven by network topology and nodes description, whereas dissemination by content characteristics.

\cite{Ashraf2010} proposes a distributed approach to routing, where each node computes the optimal route from itself to a destination broker. Each nodes knows the distance of its neighbours to the destination broker, and to break ties, the semantic distance between the node profile and the profile of its neighbours is used. A node profile instantiates a dedicated ontology, and it is serialized into a binary string. Each node profile is unique, and used as an identifier. The evaluation function computes the energetic cost of the produced routing tree, and shows a clear \textbf{reduction of energy consumption}. In this approach, the semantic web principles are used to reduce energy consumption at runtime. However, the algorithm used to serialize the semantic description of the node output is not clearly described. It outputs results that are suitable for constrained nodes, but it is unclear whether it supports the full expressivity of the ontology or not. In a different, hierarchical approach, \cite{Fredj2013} describes the construction of routing tables for service provisioning based on node clustering. Servicing nodes are described with an ontology, and the are clustered by the gateway they are connected to based on a semantic similarity measure. The obtained clusters are provided to the upper-level gateway, which uses the clusters from the nodes to build a routing table, and reproduces the process recursively. \cite{Bouhafs2006} also uses a semantic description of sensor nodes to create clusters of nodes satisfying a certain query that represents the interest of the user, for instance "Nodes detecting a temperature over 50 degrees". These clusters then aggregate the content they collect, before sending it to a sink node. The authors used the number of nodes alive against time as a metrics to show the effectiveness of their approach: the economy of energy induced by the reduction of communications allows nodes to communicate longer. However, the authors do not detail how semantic technologies are used precisely, and they make the assumption that the nodes are sufficiently powerful to match their own profile against a user query.

In these three approaches, the routes to forward content from a node to another are computed based on a semantic description of nodes. However, to the best of our knowledge, no contribution was made toward the integration of knowledge about the content itself in the routing process.

\paragraph{Querying} is the process where an application explicitly accesses content on a remote node, in a \textbf{request/response manner}. 

On the \gls{iot}, content might not necessarily be enriched and stored in a knowledge base, preventing it from being queried using semantic web technologies. \cite{Calbimonte2011} proposes a streaming \gls{obda} layer to access data in a mixed approach, where stream semantic queries are processed through R2RML mappings to be transformed into queries over federated sensor networks, and to transform the response in order to answer the original SPARQL stream query. \cite{Siow2016} proposes a similar approach, with transformation from SPARQL to SQL via RML, with and without the streaming extension to SPARQL. In other papers, where querying is not the core contribution, direct querying in SPARQL is also used after storing RDF content in a knowledge base, as it is proposed in \cite{Pfisterer2011} for instance. \cite{Poslad2015} proposes a mixed approach, where part of the data is stored in a traditionnal SQL database, but its associated metadata is stored as RDF for more expressive SPARQL queries.

The request/response interrogation of RDF graphs using SPARQL queries is\textbf{ not specific to the \gls{iot}}, and that is why few contributions in the survey cover it. This is also a reason why none of the contributions for this process can be described precisely with \gls{lmu}: the focus is not on the \gls{iot} architecture. Transformation from SPARQL to SQL is not the main practice to query \gls{iot} content, contrary to what this survey may suggest at a first glance: it is a research topic receiving contributions, as it is presented in section \ref{subsubs:obda}, dedicated to the evolutions of the semantic web.

\subsubsection{Processing content}

Processing is an important part of the \gls{iot} value creation, as is described in \cite{Qin2016}. It covers a broad range of processes (we identified 5) where content is leveraged in an application, and semantic web principles and technologies can be used in order to provide innovative services.

\paragraph{Abstraction} has similarities with \gls{cep}, as \cite{ZangLi55} points out: low level symptoms are extracted from content and correlated together in order to be transformed into a more abstract diagnosis. It can be based on reasoning, rules, pattern-recognition, etc. Abstraction requires both the collection of content, and resource-intensive processing, it is therefore an upstream process. Abstraction is a process already present in early studies such as \cite{AmitSheth30} or \cite{CoryHenson115}, where an abstraction is defined as the "representation of an environment derived from sensor observation". It is a process that has been widely implemented: 11 papers in this survey are dedicated to abstraction. Each contribution will not be presented in details in this paper, but they all have common characteristics: they use content described with an ontology (ssn for \cite{Hussein2016}, or SenML for \cite{Datta2015}), and infer new content described with a domain-specific vocabulary. The production of abstract content is necessarily \textbf{driven by an applicative need}.

\cite{AmitSheth30} proposes an SWRL-based approach to infer new content from sensor observations, illustrated by the example of the deduction of road conditions from temperature and precipitations observations. In \cite{CoryHenson115}, the authors propose a logic-based approach to abstraction, and uses OWL-DL reasoning to derive the most probable abstraction from a set of observations. \cite{Perera2014_context} takes another approach to abstraction, proposing a survey on context-awareness in the \gls{iot}, and how semantic web technologies can support the modelling of context, as well as contextual reasoning. The authors also propose several step to abstraction, with a granularity that could vary according to the layers involved in the process: content transferred from \gls{ln} to \gls{mn} is processed a first time to get a primary context, and it is processed a second time for more complex treatments on larger content instances when transferred to \gls{un}. 

Inferring new knowledge from a \gls{kb}, optionally combined to a set of rules, is a logical process not limited to the \gls{swot}. Traditionally, it is perform horizontally on a powerful machine: the \gls{kb}, the reasoner and the means to insert new content in the \gls{kb} are all hosted on the same server. However, in the case of the \gls{iot}, the process is not horizontal on \gls{un}, but rather upstream, and performed on the \gls{un} side. This is related to the nature of inference in \gls{iot} applications: it is \textbf{fed by content produced by \gls{ln}}.

\paragraph{Aggregation} is a process where multiple instances of content are used in order to produce a new content instance of the same level in the \gls{dikw} hierarchy, as opposed to abstraction where the inferred content is of a nature different from the pieces of content used for inference. For instance, computing the average (or the maximum) of several values is aggregation, while using the same values to infer the occurrence of a meteorological event is abstraction. 

In \cite{Wang2012_matching}, the aggregated elements are similarity matrices used for ontology alignment. The authors tackle the heterogeneity of application fields for the \gls{iot} domain, and the diversity of ontologies it entails, by proposing an ontology alignment technique. The proposition of these authors is not designed to be run online, and therefore it does not fit the \gls{lmu} reading grid. The proposed aggregation approach, driven by a consistency measure, is applicable to any content annotated with an ontology. In \cite{DahnLe-Phuoc18} and \cite{Le-Phuoc2016}, aggregation of content is performed using operator from SPARQL, like COUNT, MIN, etc, from a customized extension of SPARQL adapted to streaming queries to perform aggregation over time or space. In these papers, aggregation is a driving mechanism for sensor mash-up. However, content aggregation is not separated from content enrichment for instance, these two processes being both covered by the term "data fusion" in \cite{DahnLe-Phuoc18}. In these papers, the aggregated content is both produced and used in \gls{un}, making aggregation a horizontal process. 

In the papers of the survey, data fusion is not represented as an upstream process, which is counter-intuitive. It is explained by the fact that all the papers relevant to the survey perform fusion operations on content \textbf{after it has been enriched}. Therefore, the content is enriched in an upstream process, but no contribution we surveyed performed the enrichment from lower to middle nodes, and then enrichment from middle to upper nodes. In this hypothetical case, fusion would be performed upstream.

\paragraph{Visualization} is the display of content in a human-readable manner. When produced, \gls{iot} content is typically a numerical value with some metadata, and visualization processes propose a visual interpretation of it instead of its raw numerical form.

For instance, studies such as \cite{AmitSheth30} or \cite{Le-Phuoc2012} use geographical metadata to display the location of content sources on a map, and to give access to these sources via a map interface. In \cite{MohammedEid7}, an additional processing is executed to represent not only the geographical location of entities, but also to extract further meaning from the content and to attribute a color code to the sensors depending on their nature (i.e. the class they instantiate class in an ontology). In \cite{Le-Phuoc2016} (an extension of \cite{Le-Phuoc2012}), content is aggregated prior to the visualization phase, in order to display refined content, like heat maps or graphs, complementary to the map information.

\paragraph{Consistency enforcement} is the process of ensuring the consistency of the content, that is to say the absence of contradictions among the facts expressed in the content, and the absence of wrong assertions (e.g. sensor measures different from the reality of the physical world). 

\cite{Perera2014_context} discusses the role of context in the identification of inconsistency issues. \cite{Charpenay2015} proposes an ontology design pattern to capture the semantics of an industrial, non-standard, tagging vocabulary defined for the project Haystack\footnote{\url{http://project-haystack.org/}} with the semantic web formalisms. The obtained ontology is then automatically populated using the instances of Haystack tags to generate the corresponding individuals. However, the transformation from a vocabulary with limited expressivity to a rich ontology can lead to inconsistencies, for instance if individuals bear two tags from the vocabulary that are disjoint in the ontology. The authors use DL reasoning to detect inconsistencies in the generated \gls{kb}, and identify modelling issues in the ontology or inconsistencies in the individual tags.

Consistency is not instantiated in details in the contributions we surveyed, even if \cite{Corcho2010} identifies consistency as a challenge for the \gls{swot}. Contradictions between several nodes can be dealt with in an aggregation process where data fusion techniques are applied, but consistency enforcement also includes the detection of logical issues in models and their instantiations. More generally, consistency enforcement is not a topic limited to the \gls{swot}, and \textbf{generic consistency mechanisms could be applied to \gls{iot} datasets}.

\paragraph{Decision support} is a process where content is used as the input of a decision-making process. Some smart city scenarios use sensor networks and \gls{iot} deployments for decision support, e.g. Dublin's Smart Energy Demand Analysis\footnote{\url{http://smartdublin.ie/smartstories/spatial-energy-demand-analysis/}}.

\cite{Poslad2015} proposes an \gls{ews} architecture enabled with semantic functionalities. Content collected by sensors is enriched and disseminated upstream, before being used by \gls{un} in order to be abstracted into high-level events. Once the events identified, the system takes operational decisions (where to send rescue, what places to evacuate...), and these operational decisions are implemented in downstream processes. \cite{Sarkar2015} proposes the modelling of policies by ontologies in order to manage the services offered by diverse nodes, dynamically created according to context. The policies capture directives for decision making.

Decision support is a process with two aspects: on the one hand, the supported decision-making process can be performed by human beings, and on the other hand it applies to autonomic systems. The former case is not specific to the \gls{iot}, but can be instantiated with a sensor network, and the system only provides guidance to a human decision maker. 

\subsection{Node-related processes}
\label{subs:node-process}
In node-related processes, the messages exchanged between nodes do not focus on the content these nodes collect or process, but on the nodes themselves. Node-related processes classification is summarized in tables \ref{lmu-table-node}. They are separated in two sub-categories: processes dedicated to awareness among nodes, and processes dedicated to node heterogeneity management.

\subsubsection{Process related to neighbourhood awareness}

In order to communicate, \gls{iot} nodes need to \textbf{be aware of the existence of each other}, and to have respective addresses to exchange messages. Furthermore, nodes may be only intermittently available, to save battery life for example, or due to failure, maintenance operations, etc. In this context, a dynamic awareness of a node's surroundings is required, and the following processes contribute to it. 

\paragraph{Discovery} is the process of gathering descriptions of the nodes that can be reached from the node host of the process. In a typical hierarchical \gls{iot} architecture, nodes of a given level are connected to multiple nodes of an inferior level, and to a few nodes of superior level (often only one). Therefore, \gls{ln} can be deployed with the address of the middle node they communicate with hard-coded, while \gls{mn} or \gls{un} need to discover underlying nodes as they come and go. That is why discovery is a downstream process. 

\cite{Ruta2016} proposes a discovery method based on google's \gls{pw}\footnote{\url{http://google.github.io/physical-web/}} and extending it to a so-called \gls{psw}. Discovery of nearby nodes is enabled by the protocols and technologies of the \gls{pw}, but the \gls{psw} extracts from the identification messages semantic annotations in order to decide whether the node is relevant or not to the client, running on a mobile device. \cite{Kiljander2014} proposes a discovery for \gls{mn} based on a semantic description issued to a central repository which address is known a priori by the issuer of the discovery request. Similarily, SPARQL querie can be issued to the \gls{mn} describing \gls{ln} characteristics in order to collect \gls{ln} identifiers. The ontology proposed in \cite{Nambi2014} is developed in order to ease node discovery by modelling contextual knowledge, such as location, domain knowledge, and policy knowledge. \cite{SuparnaDe120}, \cite{PayamBarnaghi117}, \cite{Ming2013}, \cite{Li2015} and \cite{DavidJ.Russomanno119} all propose a discovery based on a node specification sent to a repository containing nodes descriptions. 

Discovery is a widely implemented process (10 contributions in this survey), because it is related to the dynamic nature of an \gls{iot} network: it allows \textbf{nodes' representation of the graph topography to remain consistent with the reality}. Discovery is also a process that is necessary to perform other processes. Discovering the capabilities of a distant nodes is required for processes such as lowering for instance.

\paragraph{Exposition} is the process complementary to discovery, in which a node makes its own description available in order to be discoverable. However, \textbf{exposition is initiated by the node target of the discovery}, that is why it is an upstream process.

In \cite{Ruta2016}, exposition is supported by the \gls{psw}, already described for discovery. In \cite{Alaya2015}, the authors propose to enhance the registration mechanisms offered by the oneM2M standard to support a semantic description exposition. A \gls{ln} can register itself onto a \gls{mn} if it knows its adress, and the registration query contains a standard description of the \gls{ln}. The authors propose to include a semantic description, or a dereferencable URI, in the registration request. Since then, the oneM2M standard included a "semantic descriptor" resource, carrying an RDF/XML description of its parent node. It allows a node to expose its capabilities to its target when registering onto it.  Another aspect of exposition is \textbf{proxying}: after having discovered a set of \gls{ln}, a \gls{mn} can act as a proxy and perform exposition of said \gls{ln} in their stead, as \cite{Nikoli2011} proposes. The authors propose an architecture where the exposition/discovery process can be specific to a type of network or technology. This way, the process is both adapted to the constraints of the \gls{ln} and understood by the \gls{mn}, which performs an enrichment phase on the node metadata in order to redistribute them in a more generic format.

All the contributions implementing these processes can be precisely situated in \gls{lmu}, because they all specifically describe technical content with a detailed architecture. There is significantly fewer contributions to the exposition process than to the discovery process, which seems to be counter-intuitive regarding the dependency of discovery upon exposition. However, exposition is a process requiring the lowest node to be active, while in discovery, the burden is on the highest node, making it easier to implement.

\paragraph{Selection} is a process where a node decides which other node should perform a task. It is especially relevant when multiples nodes are offering similar services yet having different capabilities, characteristics, costs, etc. Node selection is a downstream process, driven by high-level policies (e.g. energy saving or time efficiency), based on criteria to compare nodes, enabling the computation of a score and ranking. The use of semantic web technologies and principle are used to define these criteria. 

 \cite{Vlacheas2013} refers to the node selection process as the assessment of relevance metrics. The authors use abstracted nodes, or \gls{vo}, and the semantic description of their characteristics (not precisely described in the paper), in order to perform a selection. The proximity criteria are defined dynamically depending on applicative requirements. This work focuses on the ability to perform a selection despite the heterogeneity of underlying objects thanks to the abstract representation of nodes. In \cite{Perera2014}, the authors identify two types of criteria for node selection: non-negotiable criteria, representing the ability of the node to provide a service, and negotiable requirements, over which selected nodes will be ranked. The filtering phase is performed using SPARQL queries representing user requirements, and the ranking phase is based on a multi-dimensional criteria aggregation. \cite{Han2014} describes a service selection process to create a composite service out of existing services. The actual node selection is performed at binding time, where services descriptions stored in a service cache by the middle nodes is used to associate the composite service with actual nodes. In \cite{Fredj2013}, node selection is a process driven by a service search initiated by the user. This search starts from a top gateway which is the entry point of the user to a building network. Node selection is performed recursively in a gateway hierarchy, based on the semantic description of services clusters. At the lowest level of the hierarchy, the final gateway, actually connected to the lower nodes, returns the actual characteristics of the nodes matching the service request. In \cite{IoannisChatzigiannakis129}, the node selection is based on an a priori prediction of the entities that would potentially match a user query: only sensors that are associated to these entities are queried. 
 
Similarly to discovery, selection is a downstream process where the highest node perform most (if not all) the computation. Node selection can be performed to initiate a notification process: in the case of upstream notification, the lowest node has to decide whether to send a content to an upper node or not. To reduce the load on the lowest node, the process can be switched into a node selection, where the upper node selects appropriate lower nodes to provide it with content. However ,in this approach, one can see that the interest is switched from the content description in notification to the node description in node selection.

\subsubsection{Processes providing homogeneity among nodes}

As stated in the introduction, heterogeneity, and the interoperability issues it brings, is one of the leading reasons for the introduction of semantic web technologies into the \gls{iot}. The three processes dedicated to homogeneity are focus on the management of this diversity.

\paragraph{Abstraction} is the process of representing a node by a virtual entity. In a context of heterogeneous nodes, abstraction aims at focusing on the modelling of \textbf{generic nodes representations to describe their characteristic in a unified way}. It allows applications to deal with a set of homogeneous nodes, breaking the vertical silos between application domains. Nodes manipulate abstracted representations for nodes of inferior level, making abstraction an upstream process. On the \gls{swot}, \textbf{ontologies are used to describe abstract nodes}.

In \cite{Pfisterer2011}, the behavioural patterns of nodes are used in order to automatically attach a description to unclassified nodes. Similarity is computed between the output of the nodes already described in the \gls{kb} and those of the newly introduced nodes, and the undescribed nodes is annotated using the description of the similar nodes. This approach suffers a cold start issue: when only a few sensors are annotated, the clustering is less efficient, and the system needs to grow in order to propose a wider variety of sensors and more representative clusters. \cite{Vlacheas2013} proposes the notion of \gls{vo} to describe how nodes can be abstracted, and describes how this approach tackles heterogeneity issues. The authors then describe how abstracted nodes can be used in other processes, such as composition or selection. These contributions are described in the corresponding paragraphs. In \cite{Mrissa2015}, physical nodes are associated to avatars, a virtual representations described in OWL. The authors identify requirements for a \gls{wot} platform, and show how their proposed avatar architecture meets these requirements, such as interoperability, reactivity, safety... Avatars are also given introspection capabilities, in order to support collaboration and service composition, described more specifically in the paragraphs associated to these processes.

Abstraction is an important process, contributed to by 14 papers in this survey. Among the processes situated in \gls{lmu}, more contributions are situated from lower to middle nodes than from middle to upper nodes. A possible explanation is that the \textbf{\gls{ln} class is the most heterogeneous}: end devices have the widest variety of functionalities. Therefore, being able to abstract their representation as soon as possible in the \gls{iot} hierarchical network makes their management easier for \gls{mn}.

\paragraph{Composition} is the process of associating nodes between them in order to create new nodes, offering services that were unavailable on the network before. It is a process often connected to node abstraction, because an abstracted node representation enables the creation of virtual composite nodes. 

The notion of \gls{vo} presented in \cite{Vlacheas2013} and in \cite{Foteinos2013} is associated to the notion of \gls{cvo}. The authors of these contribution describe a process to \textbf{build \gls{cvo}s on top of homogeneous \gls{vo}s}, themselves being abstractions for real-world objects. Based on a specification of applicative needs, \gls{cvo}s are dynamically created.  In \cite{Han2014}, node composition is performed using the semantic description of different services. A plan is then computed, and services selected are called sequentially according to the plan. The process is dynamic, and the composite nodes described by the execution plan is not stored in the \gls{kb}. \cite{Nambi2014} uses the same terminology of service offered by a \gls{cvo} composed of several \gls{vo}. They propose a modular ontology architecture, and associate the ontology modules with the entities they define: only the \gls{vo} are described using the Resource ontology module, but it is the \gls{cvo} and the services that are annotated with the Service ontology module for instance. However, the authors do not give insight regarding the creation of the \gls{cvo}. 

Composing nodes is only possible provided that these nodes are described by abstractions, and that these abstractions can be manipulated. This is why a majority of papers are present in both abstraction and composition process: they first define how they abstract physical nodes, and then demonstrate how these abstractions can be manipulated separately from the nodes they initially represent. The semantic web principle and technologies provide flexible models that support such approach.

\paragraph{Specification/Configuration} is the process opposite to abstraction: generic representations are adapted to specific deployments, and physical nodes are configured in order to match their virtual representation.\cite{IoannisChatzigiannakis129} proposes a different definition of configuration, focused on node introspection and pattern learning: new nodes cluster with existing nodes in order to automatically annotate the data they produce. It does not qualify as configuration as it is defined in the present paper.

\cite{Kiljander2014} proposes an architecture where smart agents consume semantic messages from a broker, and control a actuators by changing their virtual representation. This process is different from control as defined in the content-centric process because the focus is not on the message that is sent in order to change the state of the device, but rather on the representation by the system of the state of the device as it should be. However, the paper does not provide further details on the implementation.

Globally, specification process as the lowering of a generic, virtual representation into potentially several implementation-specific commands has not been implemented, to the best of our knowledge.

\subsection{Identified trends}

\input{tables/table_content_1-2}

\input{tables/table_content_2-2}

\input{tables/table_node}


Four architectural characteristics emerge from existing studies: 
\begin{itemize}
    \label{item:arch-tendencies}
	\item \textit{(i)} Communication between nodes of different levels are limited to adjacent layers. Processes being supported by message flows, processes instantiations are therefore specifically represented in the tables \ref{lmu-table-content_1-2}, \ref{lmu-table-content_2-2} and \ref{lmu-table-node} between nodes of adjacent classes, e.g. from upper to middle node. 
	\item \textit{(ii)} A node of a given level has contacts with a limited number of higher level nodes, and multiple contacts with lower level nodes. 
	\item \textit{(iii)} Horizontal contacts between nodes of the same level multiply as the level of said nodes gets higher: there is little to no direct contact between lower nodes, and upper nodes communicate with each other frequently. The five contribution to horizontal processes presented in the tables are all instantiated between \gls{un}.
	\item \textit{(iv)} The \gls{iot} being a system tightly coupled with the physical world, nodes differ by their "proximity" with the environment. For a given node, this characteristic can be measured by the number of edges that separate it from a node having a direct contact (sensing or acting) in the environment. As one would expect, the lower a node is in the \gls{lmu} hierarchy, the closer it is to the environment: \gls{ln} directly observe or act on the environment, while \gls{un} only manage a digital representation of the physical world.
\end{itemize}

Situating \gls{swot} contributions in \gls{lmu} shows how \textbf{nodes constraint the processes they interact with}: contributions dedicated to the same process but at different levels in \gls{lmu} do not expect similar outcomes. However, it is worth noting that some contributions are only partially situated in \gls{lmu} because the papers describing these contributions do not provide enough implementation details. This either reveals that the deployment into a concrete architecture is not in the scope of the paper, or that the process presented is not specific to the \gls{iot} but rather uses the \gls{iot} as an application domain, for instance in the case of the contributions to the querying process. The constraints brought by the nodes justifies the presence of \cite{Kiljander2014} twice for the discovery process: discovery from upper to middle nodes being differs from discovery between middle and lower nodes. 

The superior capabilities of nodes of higher level allows system designers to \textbf{push the constraints brought by the limitations of lower nodes up}. This is why, regardless of the direction of the process, the highest nodes between the target and the source performs the more computation for content transformation and processing processes, as well as for node-centered processes. In both cases of discovery and enrichment, which are opposed processes, this observation is verified: \gls{un} are responsible for the discovery of \gls{mn} for instance, and content produced by \gls{ln} is enriched by \gls{mn}. Moreover, architectural characteristic \textit{(ii)} is a consequence of this shift of constraints: a middle node is able to manage the processing required for content enrichment that multiple lower nodes could not perform.

The influence of nodes on processes is not limited to physical constraints: the \textbf{highest nodes also holds the largest part of the application logic}. Having \gls{ln} collecting content without a specific applicative goal limits vertical fracturation: only the nodes processing content should be application-oriented. According to the architectural characteristic \textit{(iv)}, reducing vertical fracturation requires (among other measures) to reduce the application logic executed on nodes with a high priority with the environment. 

A side effect of this is the predominance of upstream processes compared to downstream processes. When comparing enrichment to lowering, or abstraction to specialization, it is clear that \textbf{upstream processes are studied more actively}. An exception to this predominance is the case of discovery and exposition, clearly visible in table \ref{lmu-table-node}: the downstream process is more represented. This is coherent with the location in the highest nodes of the application logic and the heaviest processing.  Our analysis of this matter of fact is the \textbf{similarity between upstream processes in the specific case of the \gls{iot}, and traditional research in the semantic web domain}: enriching content with metadata, or inferring abstraction from knowledge base are well-established topics. On the other hand, the need for downstream processes such as lowering of specialization arises from the constrained nature of some \gls{iot} nodes, and therefore the application domain is much more specific than the one of upstream processes. The predominance of upstream processes is coherent with the predominance of sensor nodes among \gls{iot} nodes: \textbf{downstream processes are required for actionable nodes}, which are a minority. The prevalence of upstream processes also relates to the need for homogeneity: when performing upstream operations such as node or content abstraction, the result is expressed in a rich model, providing an interoperable representation.

Moreover, in some surveyed architectures, centralization is an issue. Even if the \gls{iot} is by nature a distributed system, given its hierarchical topology, upstream processes concentrate high-level content in a few \gls{un}. When services (such as visualization) are offered directly by the server to its clients, this model is suitable, but when the network needs the capacity to make \textbf{autonomous local decisions}, this approach shows shortcomings. With the multiplication of nodes, and the increase data flow it involves, scalability is an issue for resource-intensive processes, and the next section surveys approaches to tackle this issue. Moreover, the lack of downstream processes limits the ability of \gls{mn} to benefit from the content inferred by the \gls{un} that would enable them to make local decisions. \newline

This section described contributions of the semantic web to the \gls{iot}, with the identification of generic processes and the situation of the research contributions toward these processes within \gls{lmu}. Processes such as enrichment or abstraction exist in the semantic web outside of the \gls{iot}, and they are used to solve issues specific to the \gls{iot}. However, processes such as lowering or control are more specific to the convergence between the \gls{iot} and the semantic web, and are part of evolutions of the semantic web driven by \gls{iot} constraints. These evolutions are described in the next section. 

\section{How the semantic web evolves to face IoT constraints}
\label{sec:transformation}

The issues tackled by emergence of the \gls{swot} are issues affecting the development of the \gls{iot} and the deployment of processes within \gls{iot} architectures. These issues (e.g. heterogeneity, lack of interoperability, content transformation) are recurrent concerns for the semantic web community, not necessarily related to the \gls{iot} or the \gls{wot}. However, the \gls{iot} also has intrinsic characteristics due to the constraints on its constituting nodes, the distributed nature of its deployments, and the dynamism of its topology. These constraints apply to any solution deployed in an \gls{iot} architecture. That is why \textbf{contributions between the semantic web and the \gls{iot} domain are not unidirectional}: the semantic web does contribute to the emergence of the \gls{swot}, but semantic web technologies and principles must also be adapted to meet the constraints of the \gls{iot} in order to develop the \gls{swot}.

\subsection{Adapting to the constraints on the nodes}

One of the most important constraint in the \gls{iot} is the presence of constrained physical nodes, be it \gls{ln} or, to some extent, \gls{mn}. The constraints on these nodes were presented in section \ref{subsub:node}. Among the resources that can be limited for a nodes, we distinguished energy, processing power, communication channels, and memory. These resources are not independent, e.g. the limitation of processing power or of time exchanging messages over the communication channel saves energy. In order to be suitable for an \gls{iot} network, a solution should be adaptative regarding the node running it, and the integration of semantic web principles and technologies must be thought differently on each level of \gls{lmu}: a \gls{ln} does not have the capability to run the full semantic web stack. 

Energy is a primary concern in the \gls{iot} domain, for two reasons. On the one hand, the dissemination of nodes in the environment makes it hard to connect them to a power grid (especially in non-urban area, such as fields or forests). Therefore, some nodes run on batteries, and their lifetime is directly related to their consumption of energy. On the other hand, the multiplication of nodes implies a multiplication of power consumers, and managing energy consumption at the scale of the node leads to energy saving at a more global scale. The use of semantic web technologies to reduce the cost of content transport was already discussed in the description of the associated processes presented by \cite{Bouhafs2006} and \cite{Ashraf2010}. However, in these papers, the use of semantic web technologies is not modified to match the constraints of an \gls{iot} network. 

To show how semantic web technologies can be adapted to \gls{iot} networks requirements, \cite{Su2015} compares the different serialization formats available for semantically rich data with respect to the size of the messages encoded in each format, as well as the number of CPU cycle required to produce these messages. The energy consumption associated to both the creation, the transmission, the reception and the decoding of these messages is then compared for each format. In this paper, \textbf{two aspects of energetic consumption} are pointed out: the \textbf{processing} required to handle the content, and the \textbf{communication} channels required to exchange it.

Reducing the cost of the communication can be achieved by using protocols adapted to the needs and constraints of the \gls{iot}, and by adapting existing platforms to these protocols. For instance, \cite{Loseto2016} proposes an extension of the \gls{ldp}\footnote{\url{https://www.w3.org/TR/ldp/}} specification, a recommendation of the W3C. The W3C links primitives of \gls{ldp} to HTTP\footnote{\url{https://tools.ietf.org/html/rfc2616}}, a protocol rooted in TCP\footnote{\url{https://tools.ietf.org/html/rfc793}}, therefore requiring connection between the communicating entities. HTTP is therefore not adapted to \gls{iot} architectures, where more lightweight protocols are preferred. For instance, the authors of this paper propose a mapping of \gls{ldp} to CoAP\footnote{\url{http://coap.technology/}}, based on UDP\footnote{\url{https://www.ietf.org/rfc/rfc768}}. CoAP is a protocol especially designed for constrained applications, with reduced headers and limiter packet body. Such initiative allows \gls{iot} nodes to be connected to the \gls{lod}, and therefore to extend the \gls{wot}, while respecting the constraints of \gls{iot} nodes.

Distributing content via adapted protocol also requires said content to be stored on the device distributing it. Contributions such as \cite{Bazoobandi2015} aim at allowing devices with limited memory to store semantically rich data. In this paper, the authors propose a method to store compressed RDF data in memory while ensuring a certain level of efficiency regarding encoding and decoding of data. This paper is not specifically targeted to the \gls{iot} domain, but its contribution matches the requirements of this domain. The ability to be able to access and modify efficiently the stored data is important in the case of streaming data, the authors point out. In the case of \gls{iot} deployments, and especially of sensor networks, the ability of storing and distributing up-to-date data is an important feature, and there must be a trade-off between memory optimization, and cost of encoding/decoding. \cite{Hasemann2012} also proposes a tuple store suitable for embedded systems, as well as a protocol-independant RDF broker, that can be mapped to CoAP for instance. The authors propose the adoption of a protocol stack adapted to constrained nodes in order to include them into the \gls{swot} without the need for smart gateways acting as proxies.

\begin{figure}
\centering
\includegraphics[width=0.3\textwidth]{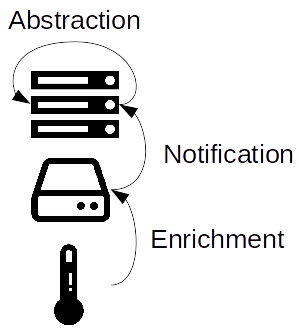}
\caption{Example of sequential processes}
\label{fig:spreading_processes}
\end{figure}

Another adaptation to the constraints of the nodes is the \textbf{splitting of an application into processes}, and the spreading of these processes across the \gls{iot} depending of the nodes able to support it. This approach is at the core of the \gls{lmu} architecture, and explains why some papers appear in the tables at different levels or for different contributions, such as \cite{Poslad2015} where the authors propose contributions to both aggregation and decision support. Figure \ref{fig:spreading_processes} gives a very simple illustration of the sequential spreading of processes, similar to \cite{Desai2015} for instance. The content generated by the \gls{ln} is enriched by the \gls{mn}, and abstracted according to applicative needs by the \gls{un} after it has been notified. 

\subsection{Managing characteristics of \gls{iot} content}

Besides the constraints on nodes, the \gls{iot} also produces content that has specific intrinsic characteristics. The usage of semantic web principles have to be adapted to this type of content as well.

\subsubsection{Stream processing}
\label{subsubs:stream}

The content dealt with on the \gls{swot} is diverse. The two families of processes identified in section \ref{subsubs:processes} (node-related and content-related) are separated by the nature of the focus of the messages they convey, and the dynamism of the content these processes rely on varies. Node-related content has both a static and a dynamic part: endpoint or API description for instance do not vary often for a given node, but its battery level or its availability can be updated frequently. Content collected by \gls{iot} nodes, on the other hand, is associated to a time and a place of collection, and is by nature dynamic. \textbf{Applications relying on such up-to-date content do not need large historic storage}.

This behaviour is associated to \textbf{stream processing}, an approach used in papers such as \cite{Calbimonte2011}, \cite{Calbimonte2012} or \cite{Le-Phuoc2016}. In these papers, part of the content, especially content describing the system itself, is static, while content generated by the system (e.g. sensor observations) is managed as an endless stream. This dual approach to content processing ensures scalability, due to the potentially infinite nature of the content stream generated by the nodes. SPARQL extensions, such as C-SPARQL\footnote{\url{http://streamreasoning.org/resources/c-sparql}} or CQELS\footnote{\url{https://github.com/cqels}}, are design to execute of continuous queries over RDF streams. Continuous querying is associated to windowing operators bounding the queried portion of the stream. \cite{Calbimonte2012} distinguishes two types of windows: time-based, containing all events over a period of time, and event-based, containing a fixed number of events. The repetitive, structured nature of \gls{iot} content makes it suitable for both windows types, the choice depending on the application.

\subsubsection{Ontology-based Data Access}
\label{subsubs:obda}
\gls{obda} is an approach to content access where the semantic web technologies are used in order to build an abstraction layer on top of a content source \cite{Kharlamov2016}. Queries are issued to the abstraction layer, and benefit from the expressivity of the semantic web technologies. The abstraction layer then transforms the query based on mappings to the destination content source language, retrieves content, and enriches content according to the mappings before returning them to the issuer of the query. This approach trades storage efficiency in large content instances for query language expressivity. Mapping languages such as R2RML\footnote{\url{https://www.w3.org/TR/r2rml/}} or more recently RML\footnote{\url{http://rml.io/}} propose a syntax to define mappings between a structured data format (relational databases for both, XML, JSON and CSV for RML).

\cite{Calbimonte2012} propose to access timestamped data streams using the SSN ontology to access streams of observations produced by sensors and stored in relational databases for the Swiss Experiment project\footnote{\url{http://www.swiss-experiment.ch}}. This approach is motivated by the heterogeneity of the underlying data models for the different data sources. \cite{Kharlamov2016} proposes an \gls{obda} approach where \textbf{static and streaming data are separated}, in order to identify clearly \textbf{background knowledge and varying observations}. Their motivation is also the homogeneity provided by the abstraction layer over the very heterogeneous underlying sensor network. The authors extend the role of the \gls{obda} layer in their system, by adding content aggregation functions to the content enrichment. \cite{Siow2016} details why \gls{iot} content is adapted to the \gls{obda} approach: the observations produced by the sensors are strongly structured, and storing only one field header for all the observations is more efficient than storing the metadata for each observation. The mapping from the metadata to the schema can be easily stored in memory, and queries on traditional relational databases are more efficient than queries on RDF stores. The limitation of this approach, however, is a reduction of the flexibility of the schema compared to an ontology, and the possibility to reason over the dataset. It is worth noting that \gls{obda} and stream processing can be complementary approaches, for instance in \cite{Calbimonte2011} and \cite{Calbimonte2012}.

Overall, this section underlines how the convergence between the \gls{iot} and the semantic web toward the \gls{swot} challenges the semantic web principles and technologies: the constraints on the nodes and specificities of \gls{iot} content are sources of innovation.

\section{Conclusion, perspectives and future work}
\label{sec:conclusion}

The \gls{iot} is a central scientific and technological topic, with a great expected societal impact. In an effort to solve interoperability issues, several \gls{iot} networks are being connected to the \gls{wot}, and semantic web principles and technologies are used to turn the \gls{wot} into the \gls{swot}. In order to be able to compare the existing \gls{iot} architectures presented by the \gls{swot} community, we proposed in this paper \gls{lmu}, a unifying \gls{iot} architectural pattern issued from a bottom-up analysis. \gls{lmu} describes a graph of nodes, where three types of nodes are separated based on their physical capabilities and their roles on the network. Generic processes have been identified through the projection of \gls{swot} contributions on \gls{lmu}, and the contributions of the semantic web to the \gls{iot} domain have been situated with respect to these processes and to \gls{lmu}. Our unified approach allowed us to identify certain trends: upstream processes are predominantly represented over downstream processes, and the capacities of the nodes influence the distribution of processes across the network.  To complement this study, evolutions of the semantic web, developed in particular to match constraints of the \gls{iot}, have been presented. This twofold approach to the landscape of the \gls{swot} showed the reciprocity of the convergence of the semantic web and the \gls{iot}: not only does the semantic web provide solutions to the interoperability and complexity issues of the \gls{iot}, but the \gls{iot} also challenges the semantic web principles and technologies to evolve to be compliant with its constraints.\newline

Even if tackled by some studies surveyed in this paper, some issues remain open for scientific contributions. For instance, \textbf{consistency} of content across nodes networks has to be ensured, in spite of content transformations when it is enriched or lowered. While contributions to content enrichment are many, table \ref{lmu-table-content_1-2} reveals a reduced number of papers in the domain of lowering, a process essential to consistency. 

The \gls{iot} is a \textbf{constrained domain}, and it requires a continuation of the research presented in section \ref{sec:transformation}. Formalisms adapted to constrained nodes and networks are emerging, proposed for instance by the W3C's \gls{wot} working group, and techniques to integrate legacy devices in semantic-aware networks need to be developed.

The increasing number of nodes is associated to an increase in the generated content volume, requiring studies such as those presented in \ref{subsubs:stream} to be extended in order to tackle \textbf{scalability} issues. The use of decentralized approaches to \gls{lmu} processes is also crucial to scalability: if sub-parts of an \gls{iot} network are able to collect, enrich, and process data in order to provide a service, the load of data exchanged on the global network is reduced, and decisions can be taken locally. Enabling local autonomic behaviours also increases resilience. However, local decisions can lead to the emergence of inconsistent global behaviours, as it is presented in the system of system theory
\cite{Lana2016}. Security, the non-functional requirement listed number one concern by open-source \gls{iot} developers in a recent survey by the Eclipse foundation\cite{eclipse2017}, also needs to be integrated deeper in the \gls{swot} in order to enable safe autonomic computing. \newline

\gls{lmu} will evolve to be adaptable to different architectural granularities, in an "unfolding" approach. In the case of very large scale deployments, such as smart cities, the node abstraction is not only suitable for devices and services, but also for self-sufficient networks. A smart building is a node compared to the city, and the capacities used for nodes classification needs to be adapted to be adapted to this recursive approach.

Furthermore, multiple perspectives are opened by the development of the \gls{swot}. Among them, a promising opportunity is the the evolution of \gls{iot} traditional machine-centric content to a richer, more expressive content, described with vocabularies connected to natural language resources. 

These resources enable the next step in semantic interoperability on the \gls{swot}, model alignment. Manual model mapping is possible in all cases, but automatic alignment requires ontologies' expressivity and machine-understandability. For instance, the FIESTA-IoT project\footnote{\url{http://fiesta-iot.eu/}} gives access to federated datasets described with a unique ontology. One the one hand, this approach ensures complete semantic interoperability, but on the other hand, it restricts the expressivity available for data providers: datasets already described with another vocabulary need to be transformed to be compliant with the fiesta ontology\footnote{\url{http://ontology.fiesta-iot.eu}}. The transformation can be manual, but it can also be systematized using automatic ontology alignments approaches. However, automatic ontologies matching techniques still need evaluation on \gls{iot} ontologies. If simple alignments are not sufficient, complex alignments should also be considered.

Moreover, being associated to \gls{m2m} communication only, \gls{iot} content is mostly numeric, and has to be interpreted by applications when displayed to a user. The enrichment of content with ontologies enables new approaches to user interaction, based on natural language for both user requirements expression, and query answering. This perspective would make \gls{iot}, \gls{wot} and \gls{swot} applications easier to access for non-expert users, bringing services to humans at the core of nodes networks.

\bibliographystyle{ieeetr}
\bibliography{SOA}
\end{document}

%% file: glossaire.tex
\newacronym{aae}{AAE}{Action-Actuator-Effect}
\newacronym{ac}{AC}{Autonomic Computing}
\newacronym{acm}{ACM}{Autonomic Computing Manager}
\newacronym{adream}{ADREAM}{Architectures dynamiques reconfigurables pour systèmes embarqués autonomes mobiles}
\newacronym{api}{API}{Application Programming Interface}
\newacronym{bc}{BC}{Base de connaissances}
\newacronym{cep}{CEP}{Complex Event Processing}
\newacronym{cps}{CPS}{Cyber-Physical System}
\newacronym{cvo}{CVO}{Composite Virtual Object}
\newacronym{dikw}{DIKW}{Data, Information, Knowledge and Wisdom}
\newacronym{dul}{DUL}{Dolce Ultra Light}
\newacronym{ews}{EWS}{Early Warning System}
\newacronym{ict}{ICT}{Information and Communication Technologies}
\newacronym{iot}{IoT}{Internet of Things}
\newacronym{iot-o}{IoT-O}{IoT Ontology}
\newacronym{irit}{IRIT}{Institut de Recherche en Informatique Toulousain}
\newacronym{kb}{KB}{Knowledge base}
\newacronym{laas}{LAAS}{Laboratoire d'Analyse et d'Architecture des Systèmes}
\newacronym{ldp}{LDP}{Linked Data Platform}
\newacronym{lmu}{LMU-N}{Lower, Middle and Upper Node}
\newacronym{ln}{LN}{Lower Node}
\newacronym{lod}{LOD}{Linked Open Data}
\newacronym{lov}{LOV}{Linked Open Vocabularies}
\newacronym{lov4iot}{LOV4IoT}{Linked Open Vocabularies for the IoT}
\newacronym{m2m}{M2M}{Machine-to-Machine}
\newacronym{mas}{MAS}{Management, Abstraction, Semantics}
\newacronym{melodi}{MELODI}{MEthodes et ingénierie des Langues, des Ontologies et du DIscours}
\newacronym{mn}{MN}{Middle Node}
\newacronym{msm}{MSM}{Minimal Service Model}
\newacronym{ndn}{NDN}{Named Data Network}
\newacronym{obda}{OBDA}{Ontology-Based Data Access}
\newacronym{odp}{ODP}{Ontology Design Pattern}
\newacronym{ows}{OWS}{Object With States}
\newacronym{owl}{OWL}{Web Ontology Language}
\newacronym{owl-s}{OWL-S}{OWL-Service}
\newacronym{pw}{PW}{Physical Web}
\newacronym{psw}{PSW}{Physical Semantic Web}
\newacronym{qudt}{QUDT}{Quantities, Units, Dimensions and Types}
\newacronym{rdf}{RDF}{Ressource Description Framework}
\newacronym{rest}{REST}{Representationnal State Transfer}
\newacronym{rpc}{RPC}{Remote Procedure Call}
\newacronym{san}{SAN}{Semantic Actuator Network}
\newacronym{sara}{SARA}{Services et Architectures pour Réseaux Avancés}
\newacronym{saref}{SAREF}{Smart Appliance REFerence}
\newacronym{sdn}{SDN}{Software Defined Network}
\newacronym{soa}{SOA}{Service Oriented Architecture}
\newacronym{sparql}{SPARQL}{SPARQL Protocol and RDF Query Language}
\newacronym{ssn}{SSN}{Semantic Sensor Network}
\newacronym{sso}{SSO}{Stimulus Sensor Observation}
\newacronym{sumo}{SUMO}{Suggested Upper Merged Ontology}
\newacronym{sweet}{SWEET}{Semantic Web for Earth and Environment Terminology}
\newacronym{swot}{SWoT}{Semantic Web Of Things}
\newacronym{swrl}{SWRL}{Semantic Web Rule Language}
\newacronym{tal}{TAL}{Traitement Automatique des Langues}
\newacronym{un}{UN}{Upper Node}
\newacronym{vo}{VO}{Virtual Object}
\newacronym{w3c}{W3C}{World Wide Web Consortium}
\newacronym{wot}{WoT}{Web Of Things}
\newacronym{wsmo}{WSMO}{Web Service Modeling Ontology}
\newacronym{wsn}{WSN}{Wireless Sensor Network}

%% file: tables/table_node_caracs.tex
\begin{table*}[]
\centering
\caption{LMU nodes physical characteristics}
\label{table-node-caracs}
\begin{tabular}{c|l|l|l|l|l|}
\cline{2-6}
\multicolumn{1}{l|}{}                                         & \multicolumn{1}{c|}{\cellcolor[HTML]{9AFF99}\textbf{Energy supply}}                             & \multicolumn{1}{c|}{\cellcolor[HTML]{9AFF99}\textbf{Processing power}}         & \multicolumn{1}{c|}{\cellcolor[HTML]{9AFF99}\textbf{Communication capabilities}}                                               & \multicolumn{1}{c|}{\cellcolor[HTML]{9AFF99}\textbf{Memory}}              & \multicolumn{1}{c|}{\cellcolor[HTML]{9AFF99}\textbf{Storage}}                               \\ \hline
\multicolumn{1}{|c|}{\cellcolor[HTML]{67FD9A}\textbf{Upper}}  & \begin{tabular}[c]{@{}l@{}}Traditional power\\ grids\end{tabular}                               & High to very high                                                              & Standard web protocols                                                                                                         & \begin{tabular}[c]{@{}l@{}}High to very high\\  (Several Go)\end{tabular} & \begin{tabular}[c]{@{}l@{}}Large to very large \\ (internal HD to\\  disk bay)\end{tabular} \\ \hline
\multicolumn{1}{|c|}{\cellcolor[HTML]{67FD9A}\textbf{Middle}} & \begin{tabular}[c]{@{}l@{}}Mixed, dependant\\ on the deployment\end{tabular}                    & Medium                                                                         & \begin{tabular}[c]{@{}l@{}}Extended, both ad-hoc and\\ standard web protocols \cite{Desai2015}\end{tabular}                           & \begin{tabular}[c]{@{}l@{}}Medium to low\\  (Up to 1Go)\end{tabular}      & \begin{tabular}[c]{@{}l@{}}Medium to limited \\ (Internal HD to\\  SD card)\end{tabular}    \\ \hline
\multicolumn{1}{|c|}{\cellcolor[HTML]{67FD9A}\textbf{Lower}}  & \begin{tabular}[c]{@{}l@{}}Often limited: battery, \\ renewable source \cite{Zanella2014}\end{tabular} & \begin{tabular}[c]{@{}l@{}}Very limited,\\  often microcontroller\end{tabular} & \begin{tabular}[c]{@{}l@{}}Constrained, Ad-hoc, potentially\\  short range (BLE, Z-Wave)\\  \cite{Desai2015} \cite{Zanella2014}\end{tabular} & \begin{tabular}[c]{@{}l@{}}Very low \\ (Under 500Ko)\end{tabular}         & \begin{tabular}[c]{@{}l@{}}Limited (Flash\\ memory, SD card)\end{tabular}                   \\ \hline
\end{tabular}
\end{table*}

%% file: tables/table_node_roles.tex
\begin{table}[]
\centering
\caption{LMU nodes roles and examples}
\label{table-node-roles}
\begin{tabular}{c|l|l|}
\cline{2-3}
\multicolumn{1}{l|}{}                                         & \multicolumn{1}{c|}{\cellcolor[HTML]{67FD9A}\textbf{Role in the network}}                                               & \multicolumn{1}{c|}{\cellcolor[HTML]{67FD9A}\textbf{Example}}                 \\ \hline
\multicolumn{1}{|c|}{\cellcolor[HTML]{67FD9A}\textbf{Upper}}  & \begin{tabular}[c]{@{}l@{}}Content processing, decision\\ making, and user interface\\ (display or API)\end{tabular}    & \begin{tabular}[c]{@{}l@{}}Cloud server,\\  laptop, tablet\end{tabular}       \\ \hline
\multicolumn{1}{|c|}{\cellcolor[HTML]{67FD9A}\textbf{Middle}} & \begin{tabular}[c]{@{}l@{}}Local content aggregation,\\  adaptation of ad-hoc com-\\ munication to the web\end{tabular} & \begin{tabular}[c]{@{}l@{}}Domestic box,\\  gateway,\\  micro-pc\end{tabular} \\ \hline
\multicolumn{1}{|c|}{\cellcolor[HTML]{67FD9A}\textbf{Lower}}  & \begin{tabular}[c]{@{}l@{}}Link to the physical world :\\ measurement and action\end{tabular}                           & Sensor, actuator                                                              \\ \hline
\end{tabular}
\end{table}

%% file: tables/table_content_1-2.tex
\begin{sidewaystable}[]
\caption{: Papers contributing to LMU content-centric processes (1/2)}
\label{lmu-table-content_1-2}
\begin{tabular}{|
>{\columncolor[HTML]{9AFF99}}c |
>{\columncolor[HTML]{9AFF99}}c |
>{\columncolor[HTML]{9AFF99}}c |l|l|l|l|l|l|}
\hline
\cellcolor[HTML]{9AFF99}                                      & \cellcolor[HTML]{9AFF99}                                & \cellcolor[HTML]{9AFF99}                              & \multicolumn{2}{c|}{\cellcolor[HTML]{9AFF99}\textbf{\begin{tabular}[c]{@{}c@{}}Representation\\  transformation\end{tabular}}}                                                                                  & \multicolumn{4}{c|}{\cellcolor[HTML]{9AFF99}\textbf{\begin{tabular}[c]{@{}c@{}}Transport/\\ Provisionning\end{tabular}}}                                                                                                                                                                                                     \\ \cline{4-9} 
\multirow{-2}{*}{\cellcolor[HTML]{9AFF99}}                    & \multirow{-2}{*}{\cellcolor[HTML]{9AFF99}\textbf{From}} & \multirow{-2}{*}{\cellcolor[HTML]{9AFF99}\textbf{To}} & \multicolumn{1}{c|}{\cellcolor[HTML]{9AFF99}\textbf{Enrichment}}                                                                               & \multicolumn{1}{c|}{\cellcolor[HTML]{9AFF99}\textbf{Lowering}} & \multicolumn{1}{c|}{\cellcolor[HTML]{9AFF99}\textbf{\begin{tabular}[c]{@{}c@{}}Notification/\\ Dissemination\end{tabular}}} & \multicolumn{1}{c|}{\cellcolor[HTML]{9AFF99}\textbf{Control}} & \multicolumn{1}{c|}{\cellcolor[HTML]{9AFF99}\textbf{Routing}} & \multicolumn{1}{c|}{\cellcolor[HTML]{9AFF99}\textbf{Querying}} \\ \hline
\cellcolor[HTML]{9AFF99}                                      & \textbf{L}                                              & \textbf{M}                                            & \cite{PayamBarnaghi118}, \cite{Desai2015}, \cite{Kiljander2014}, \cite{Maarala2014}                                                                                        &                                                                &                                                                                                                             &                                                               &                                                               &                                                                \\ \cline{2-9} 
\cellcolor[HTML]{9AFF99}                                      & \textbf{M}                                              & \textbf{U}                                            & \cite{DahnLe-Phuoc18}, \cite{Aissaoui2016}, \cite{Seydoux2016}                                                                                                      &                                                                &                                                                                                                             &                                                               &                                                               &                                                                \\ \cline{2-9} 
\multirow{-3}{*}{\cellcolor[HTML]{9AFF99}\textbf{Upstream}}   & \multicolumn{2}{c|}{\cellcolor[HTML]{9AFF99}\textbf{Unspecified}}                                               & \begin{tabular}[c]{@{}l@{}}\cite{AmitSheth30}, \cite{Ming2013}, \cite{Ganz2015}, \cite{Le-Phuoc2012}, \cite{Le-Phuoc2016}, \\ \cite{Pfisterer2011}, \cite{Compton2009}, \cite{Kotis2012a}\end{tabular} &                                                                & \cite{Qin2015}, \cite{Le-Phuoc2012}, \cite{Le-Phuoc2016}                                                                                         &                                                               &                                                               &                                                                \\ \hline
\cellcolor[HTML]{9AFF99}                                      & \textbf{U}                                              & \textbf{M}                                            &                                                                                                                                                & \cite{Seydoux2016}                                                    &                                                                                                                             & \cite{Aissaoui2016}, \cite{Seydoux2016}                                     &                                                               &                                                                \\ \cline{2-9} 
\cellcolor[HTML]{9AFF99}                                      & \textbf{M}                                              & \textbf{L}                                            &                                                                                                                                                & \cite{Kiljander2014}                                                  &                                                                                                                             &                                                               & \cite{Fredj2013}                                                     &                                                                \\ \cline{2-9} 
\multirow{-3}{*}{\cellcolor[HTML]{9AFF99}\textbf{Downstream}} & \multicolumn{2}{c|}{\cellcolor[HTML]{9AFF99}\textbf{Unspecified}}                                               &                                                                                                                                                & \cite{IoannisChatzigiannakis129}, \cite{Wang2012_ontology}, \cite{Kotis2012a}      &                                                                                                                             & \cite{IoannisChatzigiannakis129}                                     &                                                               & \cite{Qin2016}, \cite{Le-Phuoc2012}                                          \\ \hline
\cellcolor[HTML]{9AFF99}                                      & \multicolumn{2}{c|}{\cellcolor[HTML]{9AFF99}\textbf{U}}                                                         &                                                                                                                                                &                                                                &                                                                                                                             &                                                               &                                                               &                                                                \\ \cline{2-9} 
\cellcolor[HTML]{9AFF99}                                      & \multicolumn{2}{c|}{\cellcolor[HTML]{9AFF99}\textbf{M}}                                                         &                                                                                                                                                &                                                                &                                                                                                                             &                                                               &                                                               &                                                                \\ \cline{2-9} 
\multirow{-3}{*}{\cellcolor[HTML]{9AFF99}\textbf{Horizontal}} & \multicolumn{2}{c|}{\cellcolor[HTML]{9AFF99}\textbf{L}}                                                         &                                                                                                                                                &                                                                &                                                                                                                             &                                                               &                                                               &                                                                \\ \hline
\multicolumn{3}{|c|}{\cellcolor[HTML]{9AFF99}\textbf{Mixed}}                                                                                                                    &                                                                                                                                                &                                                                & \cite{Poslad2015}                                                                                                                  &                                                               & \cite{Ashraf2010}                                                    &                                                                \\ \hline
\multicolumn{3}{|c|}{\cellcolor[HTML]{9AFF99}\textbf{Non-LMU/Non specified}}                                                                                                    & \cite{Calbimonte2011}, \cite{Charpenay2015}, \cite{Ganz2014}                                                                                                        & \cite{Charpenay2015}                                                  &                                                                                                                             &                                                               & \cite{Bouhafs2006}                                                   & \cite{Calbimonte2011}, \cite{Bouhafs2006}, \cite{Siow2016}                          \\ \hline
\end{tabular}
\end{sidewaystable}

%% file: tables/table_content_2-2.tex
\begin{sidewaystable}[]
\caption{Papers contributing to LMU content-centric processes (2/2)}
\label{lmu-table-content_2-2}
\begin{tabular}{|
>{\columncolor[HTML]{9AFF99}}c |
>{\columncolor[HTML]{9AFF99}}c |
>{\columncolor[HTML]{9AFF99}}c |l|l|l|l|l|}
\hline
\cellcolor[HTML]{9AFF99}                                      & \cellcolor[HTML]{9AFF99}                                & \cellcolor[HTML]{9AFF99}                              & \multicolumn{5}{c|}{\cellcolor[HTML]{9AFF99}\textbf{Processing}}                                                                                                                                                                                                                                                                                                     \\ \cline{4-8}
\multirow{-2}{*}{\cellcolor[HTML]{9AFF99}}                    & \multirow{-2}{*}{\cellcolor[HTML]{9AFF99}\textbf{From}} & \multirow{-2}{*}{\cellcolor[HTML]{9AFF99}\textbf{To}} & \multicolumn{1}{c|}{\cellcolor[HTML]{9AFF99}\textbf{Abstraction}} & \multicolumn{1}{c|}{\cellcolor[HTML]{9AFF99}\textbf{Consistency enforcement}} & \multicolumn{1}{c|}{\cellcolor[HTML]{9AFF99}\textbf{Aggregation}} & \multicolumn{1}{c|}{\cellcolor[HTML]{9AFF99}\textbf{Visualization}} & \multicolumn{1}{c|}{\cellcolor[HTML]{9AFF99}\textbf{Decision support}} \\ \hline
\cellcolor[HTML]{9AFF99}                                      & \textbf{L}                                              & \textbf{M}                                            & \cite{Perera2014_context}                                               &                                                                               &                                                                   &                                                                     &                                                                        \\ \cline{2-8} 
\cellcolor[HTML]{9AFF99}                                      & \textbf{M}                                              & \textbf{U}                                            & \cite{Perera2014_context}, \cite{Aissaoui2016}, \cite{Datta2015}, \cite{Maarala2014}, \cite{Hussein2016}         &                                                                               &                                                                   &                                                                     &                                                                        \\ \cline{2-8}
\multirow{-3}{*}{\cellcolor[HTML]{9AFF99}\textbf{Upstream}}   & \multicolumn{2}{c|}{\cellcolor[HTML]{9AFF99}\textbf{Unspecified}}                                                       & \cite{ZangLi55},\cite{PayamBarnaghi117}, \cite{Qin2016}                           &                                                            &                                    & \cite{PayamBarnaghi117}, \cite{Le-Phuoc2012} , \cite{Le-Phuoc2016}                                     &                                                                        \\ \hline
\cellcolor[HTML]{9AFF99}                                      & \textbf{U}                                              & \textbf{M}                                            &                                                                   &                                                                               &                                                                   &                                                                     &                                                                        \\ \cline{2-8}
\cellcolor[HTML]{9AFF99}                                      & \textbf{M}                                              & \textbf{L}                                            &                                                                   &                                                                               &                                                                   &                                                                     &                                                                        \\ \cline{2-8}
\multirow{-3}{*}{\cellcolor[HTML]{9AFF99}\textbf{Downstream}} & \multicolumn{2}{c|}{\cellcolor[HTML]{9AFF99}\textbf{Unspecified}}                                                       &                                                                   &                                                                               &                                                                   &                                                                     &                                                                        \\ \hline
\cellcolor[HTML]{9AFF99}                                      & \multicolumn{2}{c|}{\cellcolor[HTML]{9AFF99}\textbf{U}}                                                         &                                        &                                                                               & \cite{Poslad2015},  \cite{DahnLe-Phuoc18}     , \cite{Le-Phuoc2016}                                  &                                                                     & \cite{Poslad2015}                                                             \\ \cline{2-8}
\cellcolor[HTML]{9AFF99}                                      & \multicolumn{2}{c|}{\cellcolor[HTML]{9AFF99}\textbf{M}}                                                         &                                                                   &                                                                               &                                                                   &                                                                     &                                                                        \\ \cline{2-8}
\multirow{-3}{*}{\cellcolor[HTML]{9AFF99}\textbf{Horizontal}} & \multicolumn{2}{c|}{\cellcolor[HTML]{9AFF99}\textbf{L}}                                                         &                                                                   &                                                                               &                                                                   &                                                                     &                                                                        \\ \hline
\multicolumn{3}{|c|}{\cellcolor[HTML]{9AFF99}\textbf{Mixed}}                                                                                                                    &                                                        &                                                                               &                                                            &                                                                     &                                                                        \\ \hline
\multicolumn{3}{|c|}{\cellcolor[HTML]{9AFF99}\textbf{Non-LMU/Non specified}}                                                                                                    & \cite{AmitSheth30}, \cite{CoryHenson115}, \cite{Ganz2014}                              & \cite{Perera2014_context}, \cite{Charpenay2015}                                                                 &        \cite{Wang2012_matching}, \cite{Kharlamov2016}       & \cite{AmitSheth30}, \cite{MohammedEid7}                                                        & \cite{Wu2014}, \cite{Sarkar2015}                                                     \\ \hline
\end{tabular}
\end{sidewaystable}

%% file: tables/table_node.tex
\begin{sidewaystable}[]
\centering
\caption{Papers contributing to LMU node-centric processes}
\label{lmu-table-node}
\begin{tabular}{|
>{\columncolor[HTML]{9AFF99}}c |
>{\columncolor[HTML]{9AFF99}}c |
>{\columncolor[HTML]{9AFF99}}l |l|l|l|l|l|l|}
\hline
\cellcolor[HTML]{9AFF99}                                      & \cellcolor[HTML]{9AFF99}                                & \cellcolor[HTML]{9AFF99}                                & \multicolumn{3}{c|}{\cellcolor[HTML]{9AFF99}\textbf{Awareness}}                                                                                                                                                                                                                                                                                    & \multicolumn{3}{c|}{\cellcolor[HTML]{9AFF99}\textbf{Homogeneity}}                                                                                                                                                                                                                                                                                                       \\ \cline{4-9} 
\multirow{-2}{*}{\cellcolor[HTML]{9AFF99}}                    & \multirow{-2}{*}{\cellcolor[HTML]{9AFF99}\textbf{From}} & \multirow{-2}{*}{\cellcolor[HTML]{9AFF99}\textbf{To}}   & \multicolumn{1}{c|}{\cellcolor[HTML]{9AFF99}\textbf{Discovery}}                                                                             & \multicolumn{1}{c|}{\cellcolor[HTML]{9AFF99}\textbf{Exposition}} & \multicolumn{1}{c|}{\cellcolor[HTML]{9AFF99}\textbf{Selection}}                                                                   & \multicolumn{1}{c|}{\cellcolor[HTML]{9AFF99}\textbf{Abstraction}}                                                                              & \multicolumn{1}{c|}{\cellcolor[HTML]{9AFF99}\textbf{Composition}}                       & \multicolumn{1}{c|}{\cellcolor[HTML]{9AFF99}\textbf{\begin{tabular}[c]{@{}c@{}}Specification\\ /Configuration\end{tabular}}} \\ \hline
\cellcolor[HTML]{9AFF99}                                      & \textbf{L}                                              & \textbf{M}                                              &                                                                                                                                             & \cite{Alaya2015}, \cite{Ruta2016}                                              &                                                                                                                                   & \begin{tabular}[c]{@{}l@{}}\cite{Kibria2015}, \cite{Nikoli2011}, \cite{Kiljander2014}, \\ \cite{Fredj2013}\end{tabular}                                                    & \cite{Kibria2015}                                                                              &                                                                                                                              \\ \cline{2-9} 
\cellcolor[HTML]{9AFF99}                                      & \textbf{M}                                              & \textbf{U}                                              &                                                                                                                                             & \cite{L2016}, \cite{Nikoli2011}                                                &                                                                                                                                   & \cite{Foteinos2013}, \cite{Aissaoui2016}                                                                                                                     & \cite{DahnLe-Phuoc18}, \cite{Foteinos2013}                                                            &                                                                                                                              \\ \cline{2-9} 
\multirow{-3}{*}{\cellcolor[HTML]{9AFF99}\textbf{Upstream}}   & \multicolumn{2}{c|}{\cellcolor[HTML]{9AFF99}\textbf{Unspecified}}                                                 &                                                                                                                                             &                                                                  &                                                                                                                                   & \begin{tabular}[c]{@{}l@{}}\cite{SuparnaDe120}, \cite{Perera2014}, \cite{Vlacheas2013}, \\ \cite{Mrissa2015}, \cite{Nambi2014}, \cite{Wang2012_ontology}, \\ \cite{Pfisterer2011}\end{tabular} & \begin{tabular}[c]{@{}l@{}}\cite{Vlacheas2013}, \cite{Mrissa2015}, \cite{Nambi2014}, \\ \cite{Han2014}\end{tabular} &                                                                                                                              \\ \hline
\cellcolor[HTML]{9AFF99}                                      & \textbf{U}                                              & \multicolumn{1}{c|}{\cellcolor[HTML]{9AFF99}\textbf{M}} & \cite{Kibria2015}, \cite{Kiljander2014}, \cite{Ruta2016}                                                                                                         &                                                                  & \cite{Han2014}                                                                                                                           &                                                                                                                                                &                                                                                         &                                                                                                                              \\ \cline{2-9} 
\cellcolor[HTML]{9AFF99}                                      & \textbf{M}                                              & \multicolumn{1}{c|}{\cellcolor[HTML]{9AFF99}\textbf{L}} & \cite{Kiljander2014}                                                                                                                               &                                                                  & \cite{Fredj2013}                                                                                                                         &                                                                                                                                                &                                                                                         & \cite{Kiljander2014}                                                                                                                \\ \cline{2-9} 
\multirow{-3}{*}{\cellcolor[HTML]{9AFF99}\textbf{Downstream}} & \multicolumn{2}{l|}{\cellcolor[HTML]{9AFF99}\textbf{Unspecified}}                                                 & \begin{tabular}[c]{@{}l@{}}\cite{SuparnaDe120}, \cite{PayamBarnaghi117}, \cite{PayamBarnaghi117},\\ \cite{Ming2013}, \cite{Nambi2014}\end{tabular} &                                                                  & \begin{tabular}[c]{@{}l@{}}\cite{DahnLe-Phuoc18}, \cite{Perera2014}, \cite{Vlacheas2013}, \\ \cite{PayamBarnaghi117}, \cite{Pfisterer2011}, \cite{Hussein2016},\\\cite{IoannisChatzigiannakis129}\end{tabular} &                                                                                                                                                &                                                                                         &                                                                                                                              \\ \hline
\cellcolor[HTML]{9AFF99}                                      & \multicolumn{2}{c|}{\cellcolor[HTML]{9AFF99}\textbf{U}}                                                           & \cite{Li2015}                                                                                                                                      &                                                                  &                                                                                                                                   &                                                                                                                                                &                                                                                         &                                                                                                                              \\ \cline{2-9} 
\cellcolor[HTML]{9AFF99}                                      & \multicolumn{2}{c|}{\cellcolor[HTML]{9AFF99}\textbf{M}}                                                           &                                                                                                                                             &                                                                  &                                                                                                                                   &                                                                                                                                                &                                                                                         &                                                                                                                              \\ \cline{2-9} 
\multirow{-3}{*}{\cellcolor[HTML]{9AFF99}\textbf{Horizontal}} & \multicolumn{2}{c|}{\cellcolor[HTML]{9AFF99}\textbf{L}}                                                           &                                                                                                                                             &                                                                  &                                                                                                                                   &                                                                                                                                                &                                                                                         &                                                                                                                              \\ \hline
\multicolumn{3}{|c|}{\cellcolor[HTML]{9AFF99}\textbf{Mixed}}                                                                                                                      &                                                                                                                                             &                                                                  &                                                                                                                                   &                                                                                                                                                &                                                                                         &                                                                                                                              \\ \hline
\multicolumn{3}{|c|}{\cellcolor[HTML]{9AFF99}\textbf{Non-LMU/Unspecified}}                                                                                                        & \cite{DavidJ.Russomanno119}                                                                                                                        &                                                                  & \cite{Bouhafs2006}                                                                                                                       & \cite{MichaelCompton27}                                                                                                                               &                                                                                         &                                                                                                                              \\ \hline
\end{tabular}

\end{sidewaystable}